\documentclass[proof]{pasj01}
\usepackage{multirow}
\usepackage{graphicx}
\Received{$\langle$reception date$\rangle$}
\Accepted{$\langle$acception date$\rangle$}
\Published{$\langle$publication date$\rangle$}

\newcommand{\fDG}{f_{\rm DG}}
\newcommand{\fthree}{f_{\rm ^{13}CO}}
\newcommand{\feight}{f_{\rm C^{18}O}}

\newcommand{\ffDG}{f^{13}_{\rm DG}}
\newcommand{\ffeight}{f^{13}_{\rm C^{18}O}}

\newcommand{\fffDG}{f^{18}_{\rm DG}}

\newcommand{\mH}{M_{\rm H_2}}
\newcommand{\mDG}{M_{\rm H_2}{\rm (DG)}}
\newcommand{\mtwo}{M_{\rm H_2}{\rm (^{12}CO)}}
\newcommand{\mthree}{M_{\rm H_2}{\rm (^{13}CO)}}
\newcommand{\meight}{M_{\rm H_2}{\rm (C^{18}O)}}

\begin{document}

\title{{ 
FOREST Unbiased Galactic Plane Imaging Survey with the Nobeyama 45-m Telescope (FUGIN) V: Dense gas mass fraction of molecular gas in the Galactic plane
}}
\author{Kazufumi Torii\altaffilmark{1}, Shinji Fujita\altaffilmark{2}, Atsushi Nishimura\altaffilmark{2}, Kazuki, Tokuda\altaffilmark{3,4}, Mikito Kohno\altaffilmark{2}, Kengo Tachihara\altaffilmark{2}, Shu-ichiro Inutsuka\altaffilmark{2}, Mitsuhiro Matsuo\altaffilmark{1}, Mika Kuriki\altaffilmark{5}, Yuya Tsuda\altaffilmark{7}, Tetsuhiro Minamidani\altaffilmark{1,8}, Tomofumi Umemoto\altaffilmark{1,8}, Nario Kuno\altaffilmark{5,6}, Yusuke Miyamoto\altaffilmark{4}}%
\altaffiltext{1}{Nobeyama Radio Observatory, 462-2 Nobeyama Minamimaki-mura, Minamisaku-gun, Nagano 384-1305, Japan}
\altaffiltext{2}{Graduate School of Science, Nagoya University, Chikusa-ku, Nagoya, Aichi 464-8601, Japan}
\altaffiltext{3}{Department of Physical Science, Graduate School of Science, Osaka Prefecture University, 1-1 Gakuen-cho, Naka-ku, Sakai, Osaka 599-8531, Japan }
\altaffiltext{4}{Chile Observatory, National Astronomical Observatory of Japan, National Institutes of Natural Science, 2-21-1 Osawa, Mitaka, Tokyo 181-8588, Japan}
\altaffiltext{5}{Department of Physics, Graduate School of Pure and Applied Sciences, University of Tsukuba, 1-1-1 Ten-nodai, tsukuba, Ibaraki 305-8577, Japan}
\altaffiltext{6}{Tomonaga Center for the History of the Universe, University of Tsukuba, Tsukuba, Ibaraki 305-8571, Japan}
\altaffiltext{7}{Meisei University, 2-1-1 Hodokubo, Hino, Tokyo 191-0042, Japan}
\altaffiltext{8}{Department of Astronomical Science, School of Physical Science, SOKENDAI (The Graduate University for Advanced Studies), 2-21-1, Osawa, Mitaka, Tokyo 181-8588, Japan}

\email{kazufumi.torii@nao.ac.jp}

\KeyWords{ISM: clouds --- ISM: molecules --- radio lines: ISM --- stars: formation}

\maketitle

\begin{abstract}
Recent observations of the nearby Galactic molecular clouds indicate that the dense gas in molecular clouds have quasi-universal properties on star formation, and observational studies of extra-galaxies have shown a galactic-scale correlation between the star formation rate (SFR) and surface density of molecular gas. 
To reach a comprehensive understanding of both properties, it is important to quantify the fractional mass of the dense gas in molecular clouds $\fDG$. In particular, for the Milky Way (MW), there are no previous studies resolving the $\fDG$ disk over a scale of several kpc.
In this study, the $\fDG$ was measured over 5\,kpc in the first quadrant of the MW, based on the CO $J$=1--0 data in $l=10^\circ$--$50^\circ$ obtained as part of the FOREST Unbiased Galactic Plane Imaging Survey with the Nobeyama 45-m Telescope (FUGIN) project. 
The total molecular mass was measured using $^{12}$CO, and the dense gas mass was estimated using C$^{18}$O.
The fractional masses including $\fDG$ in the region within $\pm 30\%$ of the distances to the tangential points of the Galactic rotation (e.g., the Galactic Bar, Far-3kpc Arm, Norma Arm, Scutum Arm, Sagittarius Arm, and inter-arm regions) were measured.
As a result, an averaged $\fDG$ of $2.9^{+2.6}_{-2.6}\%$ was obtained for the entirety of the target region.
This low value suggests that dense gas formation is the primary factor of inefficient star formation in galaxies.
It was also found that the $\fDG$ shows large variations depending on the structures in the MW disk. 
The $\fDG$ in the Galactic arms were estimated to be $\sim4$--$5\%$, while those in the bar and inter-arm regions were as small as $\sim0.1$--$0.4\%$.
These results indicate that the formation/destruction processes of the dense gas and their timescales are different for different regions in the MW, leading to the differences in SFRs.
\end{abstract}

\section{Introduction}
Star formation in galaxies are characterized by the Kennicutt-Schmidt (KS)-law \citep{sch1959, ken1998, ken2012}, which is a galactic-scale empirical correlation between the area-averaged Star Formation Rate (SFR) ($\Sigma_{\rm SFR} \, [M_\odot\,{\rm yr^{-1}\, kpc^{-2}}]$) and gas surface density ($\Sigma_{\rm H_2+H}\, [M_\odot\,{\rm pc^{-2}}]$) with a power-law index of $\sim$1.4 ($\Sigma_{\rm SFR} \propto \Sigma_{\rm H_2+H}^N$). 
This correlation can be seen in the inner-parts of the galaxies where H$_2$ is dominant (e.g., \cite{tan2014,sof2016}), indicating an index of $\sim1$ with scatters of approximately 0.2\,dex ($\Sigma_{\rm SFR} \propto \Sigma_{\rm H_2}$, \cite{big2008}).
The scatter is likely due to the regional differences in Star Formation Efficiency (SFE) in the individual galaxies (i.e., bar, arm, inter-arm, and nucleus) (e.g., \cite{mom2010}).
The KS-law predicts the gas consumption timescale of $H_2$ gas to be $\tau_{\rm con} = \Sigma_{\rm H_2}/\Sigma_{\rm SFR}$ of $\sim$1--2\,Gyr (e.g., \cite{big2011}), which is three orders of magnitude larger than a free-fall timescale of $\sim1$\,Myr at a gas density of 100\,cm$^{-3}$.
Understanding the background physics of inefficient star formation in galaxies is one of the most pressing issues in contemporary astrophysics.

While the $\Sigma_{\rm H_2}$ used in the KS-law is generally measured using the CO rotational transition emission, several observations used the HCN $J$=1--0 transition with a critical density of $\sim2\times10^6$\,cm$^{-3}$ (which is  in reality reduced by radiative trapping owing to its high optical depth, \cite{gao2004a}) to measure the mass (or luminosity) of dense molecular gas to construct a dense-gas KS-law \citep{gao2004a,gao2004b,use2015,big2016}. 
Their results indicate a tighter correlation with $\Sigma_{\rm SFR}$ rather than $\Sigma_{\rm H_2}$.

The recent submillimeter imaging surveys of Galactic molecular clouds with the {\it Herschel} Space Observatory have made remarkable progress in understanding the star formation in dense gas.
The {\it Herschel} observations demonstrate that the molecular filaments take up a dominant fraction of dense gas in molecular clouds \citep{and2010, mol2010, kon2015, arz2018a}. 
These filaments are characterized by the narrow distribution of central widths with a full width at half maximum (FWHM) of $\sim0.1$\,pc \citep{arz2011}.

An important discovery made by the {\it Herschel} observations was that the majority of prestellar cores are embedded within ``supercritical'' filaments \citep{and2014}, for which the mass per unit length exceeds the critical line mass, $M_{\rm line, crit} = 2 c_{\rm s}^2/G \sim 16\,M_\odot\,{\rm pc^{-1}}$ (e.g., \cite{inu1997}), where $c_{\rm s}\sim0.2$\,km\,s$^{-1}$ is the isothermal sound speed at temperature $T \sim 10$\,K, and $G$ is the gravitational constant.
Given the filament width of $\sim0.1$\,pc, the $M_{\rm line, crit} \sim 16\,M_\odot\,{\rm yr^{-1}}$ predicts a quasi-universal threshold for core/star formation in molecular clouds at $\Sigma_{\rm H_2} \sim 160\,M_\odot\,{\rm yr^{-1}}$ in terms of the gas surface density, which corresponds to an H$_2$ column density $N_{\rm H_2}$ of $\sim7\times 10^{21}$\,cm$^{-2}$ or a visual extinction $A_{\rm v}$ of $\sim8$\,mag 
(assuming $N_{\rm H_2}/A_{\rm v} = 0.94\times10^{21}$, \cite{boh1978}).
Such a threshold for star formation was also discussed in independent observational studies; \citet{oni1998} proposed star formation of $N_{\rm H_2} \geq 8\times10^{21}$\,cm$^{-2}$ based on the C$^{18}$O observations of the Taurus molecular cloud.
The {\it Spitzer} infrared observations of Galactic nearby clouds provided a similar threshold of $\Sigma_{\rm H_2} \sim 130\,M_\odot\,{\rm yr^{-1}}$ \citep{hei2010}.

Measurements of SFE in dense gas, or supercritical filaments, were performed on nearby Galactic molecular clouds \citep{wu2005, lad2010, lad2012, shi2017}.
The studies of \citet{lad2010} and \citet{shi2017} were done on gas with $A_{\rm v} > 8$\,mag, presenting results that were consistent with the study on extra-galaxies by \citet{gao2004a} (see also \cite{big2016}); the gas consumption timescale of the dense gas can be computed as $\tau_{\rm con}\sim20$\,Myr.
This implies that the SFE in dense molecular gas in galaxies is quasi-universal on scales from $\sim$1--10\,pc to $>$10\,kpc.
\citet{lad2010} proposed that $\Sigma_{\rm SFR} \propto f_{\rm DG}\Sigma_{\rm H_2}$, where $f_{\rm DG}$ is the mass fraction of dense gas to molecular gas, is the fundamental relationship governing star formation in galaxies.
Following these studies, this paper defines ``dense gas'' as gas with $A_{\rm v} > 8$\,mag.

For extra-galaxies, \citet{mur2016} revealed the temperature and density distribution of the molecular gas in NGC\,2903 using large velocity gradient analysis \citep{gol1974, sco1974}, presenting a positive correlation between gas densities and SFE.
HCN observations of extra-galaxies by \citet{use2015} and \citet{big2016} found that SFE in dense gases depend on galactic environment, being lower at high stellar surface densities and high H$_2$-to-H{\sc i} mass ratio.

These studies emphasize the importance of measuring $f_{\rm DG}$ in various galactic environments and understanding its relationship with star formation.
In the MW, \citet{bat2014} measured the H$_2$ mass fractions of dense gas components in Giant Molecular Clouds (GMCs) in the Galactic plane. 
The masses of the GMCs were calculated using the $^{13}$CO $J$=1--0 data for $l=18^\circ$--$56^\circ$ taken by the Five College Radio Astronomical Observatory (FCRAO), while those of the dense gas were calculated from the Bolocam Galactic Plane Survey (BGPS) 1.1\,mm dust continuum images. 
The radii and masses of their GMC samples cover $\sim5$--$20$\,pc and $10^3$--$10^5$\,$M_\odot$, respectively.
The authors obtained a low averaged fractional mass of 11$^{+12}_{-06}\,\%$, and the derived mass fractions are independent of the GMC masses.
Since the GMC masses were derived in $^{13}$CO, they cannot be directly compared to the KS-law in extra-galaxies, in which $\Sigma_{\rm H_2}$ is usually measured using $^{12}$CO.
Based on the $^{12}$CO and CS observations toward a 2\,deg$^2$ area at $l\sim44\fdg1$--$46\fdg3$, \citet{rom2016} obtained a $f_{\rm DG}$ of $\sim14\%$ over $\sim200$--$300$\,pc at $R$ of $6$--$8.5$\,kpc.
To date, the $f_{\rm DG}$ of the Galactic plane has not been derived for kpc scales.

In this study, the $^{12}$CO, $^{13}$CO, and C$^{18}$O $J$=1--0 data obtained for $l=10^\circ$--$50^\circ$ using the Nobeyama 45-m radio telescope were analyzed to measure the H$_2$ mass ($M_{\rm H_2}$) of molecular gas detected independently in the three CO isotopologues. 
The $J$=1--0 transition of CO has a critical density of $\sim2\times10^3$\,cm$^{-3}$, and the differences on the abundance ratios of the three CO isotopologues (which lead to differences on the optical depths along the line-of-sight) allow us to probe different ranges of $N_{\rm H_2}$ in the molecular clouds.

Our analyses include the Galactic bar, Far-3kpc Arm, Norma Arm, Scutum-Centaurus Arm, and Sagittarius Arm, as well as the inter-arm regions of these arms.
The mass fractions of dense molecular gas in these regions were first measured by taking the mass ratios of the $^{13}$CO and C$^{18}$O emitting gas to the $^{12}$CO emitting gas.

The $J$=1--0 transition of $^{12}$CO is known as a tracer of the total $\mH$ of molecular clouds, although it is consistently found to be optically thick. 
Comparisons between the $^{12}$CO integrated intensities $W({\rm ^{12}CO})$ with other $\mH$ tracers (e.g., virial mass, gamma-ray emission, and dust emission) indicated a surprisingly close correlation between $W({\rm ^{12}CO})$ and the $N_{\rm H_2}$ of the molecular clouds (e.g., \cite{dic1978,sand1984,sol1987,str1996,dam2001,planck2011}). 
This provides a CO-to-H$_2$ conversion factor in the inner Galaxy at the galactocentric radius $R$ of 1--9\,kpc as reported and expressed by the equation\,\ref{eq0} \citep{bol2013},
\begin{equation}
X({\rm CO}) \ = \  \frac{N_{\rm H_2}}{W({\rm ^{12}CO})} \ = \ 2 \times 10^{20} {\rm (K\,km\,s^{-1})^{-1}\,cm^{-2}}. \label{eq0}
\end{equation}
The uncertainty of the $X$(CO) was discussed to be as small as a factor of 1.3 by \citet{bol2013}.
$X$(CO) has been utilized to measure the $\Sigma_{\rm H_2}$ in other galaxies; therefore, it is important to measure the $\mH$ of the molecular gas in the MW from the $^{12}$CO $J$=1--0 data and $X$(CO) to evaluate $f_{\rm DG}$.

Although $^{12}$CO $J$=1--0 is essentially a good $\mH$ tracer of molecular clouds, it barely works as an $N_{\rm H_2}$ tracer in the dense parts of molecular clouds that have large $N_{\rm H_2}$ due to intensity saturation by the opacity effect, and therefore less abundant $^{13}$CO and C$^{18}$O are typically used to measure more accurate $\mH$ and $N_{\rm H_2}$ in these parts.
Low-$J$ transitions of $^{13}$CO have been used in the observations of nearby GMCs which have a typical density of $\sim10^3$\,cm$^{-3}$ (e.g., \cite{miz1995,nag1998,gol2008,nar2008,nis2015}).
C$^{18}$O is optically thin even in the higher $N_{\rm H_2}$ parts of the molecular clouds, which have dominant filamentary structures of a 0.1\,pc width (e.g., \cite{oni1996,oni1998,hac2013,nis2015,tok2018,arz2018a}).
Therefore, C$^{18}$O can be used to measure the $\mH$ of the dense gas to evaluate $f_{\rm DG}$ in the MW.

Here, it is noteworthy that CO molecules sometimes do not trace $N_{\rm H_2}$ in dense cores at the innermost parts of molecular clouds owing to the heavy opacities and/or molecular depletion of dust grains (e.g., \cite{cas1999,ber2007}). 
However, as the Herschel observations revealed, the ratio of the mass in the dense star forming cores to the mass of the parental filament is less than $\sim15$\% on average \citep{and2014,kon2015}. This means that possible depletion of molecules in the dense regions has only a limited effect on the estimation of the total mass in the filaments. Likewise the molecular depletion is not expected to be significant in our mass estimate from the C$^{18}$O data with pc-scale resolution in this study.

The remainder of this paper is organized as follows.
Section\,2 describes the CO $J$=1--0 dataset used in this study. 
Section\,3 presents the target region of the current analyses.
Section\,4 presents the main results of analyzing the CO dataset. 
Results are discussed in Section\,5, and, finally, a summary is presented in Section\,6.

\section{Dataset}
The $^{12}$CO, $^{13}$CO, and C$^{18}$O $J$=1--0 datasets obtained by the FOREST Unbiased Galactic Plane Imaging survey using the Nobeyama 45-m telescope (FUGIN; see \cite{ume2017} for a full description of the observations and data reduction) were analyzed. 
FUGIN involved a large-scale Galactic plane survey using the FOur-beam REceiver System on the 45-m Telescope (FOREST; \cite{min2016}); the four-beam, dual-polarization, two-sideband receiver installed in the Nobeyama 45-m telescope. 
This study utilized the FUGIN dataset obtained at $l=10^\circ$--$50^\circ$ in the first quadrant of the Galactic plane.
The typical system temperatures were $\sim$250\,K for $^{12}$CO $J$=1--0 (115.271\,GHz) and $\sim$150\,K for $^{13}$CO $J$=1--0 (110.201\,GHz) and C$^{18}$O $J$=1--0 (109.782\,GHz).
The backend system was the digital spectrometer ``SAM45'' \citep{kun2011, kam2012}, which provided a bandwidth of 1\,GHz and a resolution of 244.14\,kHz. 
These figures corresponded to 2,600\,km\,s$^{-1}$ and 0.65\,km\,s$^{-1}$ at 115\,GHz, respectively.  
The observations were made in the On-The-Fly mode with a unit map size of $1^\circ \times 1^\circ$.
The pointing accuracy was checked almost every hour to keep within 3$''$ by observing SiO maser sources.
The output data were formatted into a size of $1^\circ \times 2^\circ$ with spatial and velocity grid-sizes of 8.5$''$ and 0.65\,km\,s$^{-1}$, respectively. 
Absolute intensity calibrations were performed by adopting Main beam efficiencies of $0.45\pm0.02$ and $0.43\pm0.02$ at 110 and 115\,GHz, respectively \citep{ume2017}.

\begin{figure}
 \begin{center}
  \includegraphics[width=14cm]{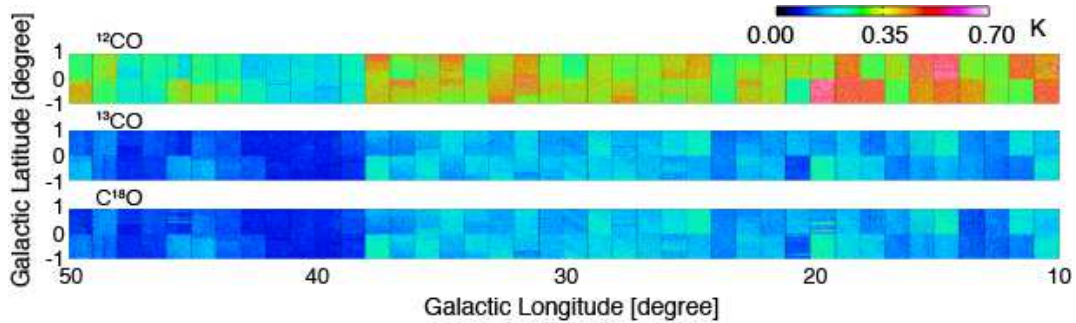}
 \end{center}
 \caption{The r.m.s noise ($\sigma$) distributions of the $^{12}$CO, $^{13}$CO, and C$^{18}$O data. }\label{fig:rms}
\end{figure}

The output CO data, particularly $^{12}$CO data, suffers from scan-effects and spurious-like structures.
To remove these features as well as to improve the sensitivity, the following post-processes were applied to the output data cube: 
(1) one-dimensional median filtering to the velocity axis with a kernel of 3\,ch,
(2) two-dimensional median filtering to the spatial axes with a kernel of $3\times3$\,ch,
and (3) two-dimensional spatial smoothing with a Gaussian function to achieve a spatial resolution of 40$''$. 
Figure\,\ref{fig:rms} presents the root-mean-square (r.m.s) 1\,ch noise $\sigma$ distributions of the post-processed $^{12}$CO, $^{13}$CO, and C$^{18}$O data.
Although the post-process efficiently removes the scan-effects and spurious structures, some of these features still remains in some tiles.
These remaining features were finally removed in identifying the CO sources (Section\,4.2).

\section{Region Selection}
The vertical distributions of $^{12}$CO in the inner Galaxy were measured to be $\sim50$--$100$\,pc at FWHM (e.g., \cite{nak2006}). Thus, it is necessary to cover $\sim200$\,pc in $b$ to accurately measure the total $M_{\rm H_2}$ in $^{12}$CO.
It is also important to achieve a spatial resolution of less than a few pc to detect the dense gas components in molecular clouds (e.g., \cite{ber2007}). 
Furthermore, in order to estimate $\fDG$ accurately, it is of primary importance to reduce the errors on estimating distances of molecular clouds, as the above requirements for the $b$ coverage and spatial resolution cannot be guaranteed otherwise.

Considering these conditions required to measure $f_{\rm DG}$ in the Galactic plane, the focus was placed on the tangential points of the Galactic rotation relative to the local standard of rest (LSR), at which the radial velocities of the CO emissions $v_{\rm LSR}$ correspond to the terminal radial velocities of the Galactic rotation $v_{\rm term}$, and one unique solution in kinematic distance can be given. 
The kinematic distance to the tangential points are refereed to as $d_{\rm tan}$, which depends only on $l$, from here on out.
Assuming the IAU standard parameters (the distance to the Galactic center $R_0 = 8.5$\,kpc and the LSR rotational velocity $\Theta_0 = 220$\,km\,s$^{-1}$), the $l$ coverage of the FUGIN observations ($l=10^\circ$--$50^\circ$) correspond to $d_{\rm tan}$ of $\sim5.5$--$8.4$\,kpc and the galactocentric distances to the tangential points $R_{\rm tan}$ of $\sim1.5$--$6.5$\,kpc.

\begin{figure}
 \begin{center}
  \includegraphics[width=15.5cm]{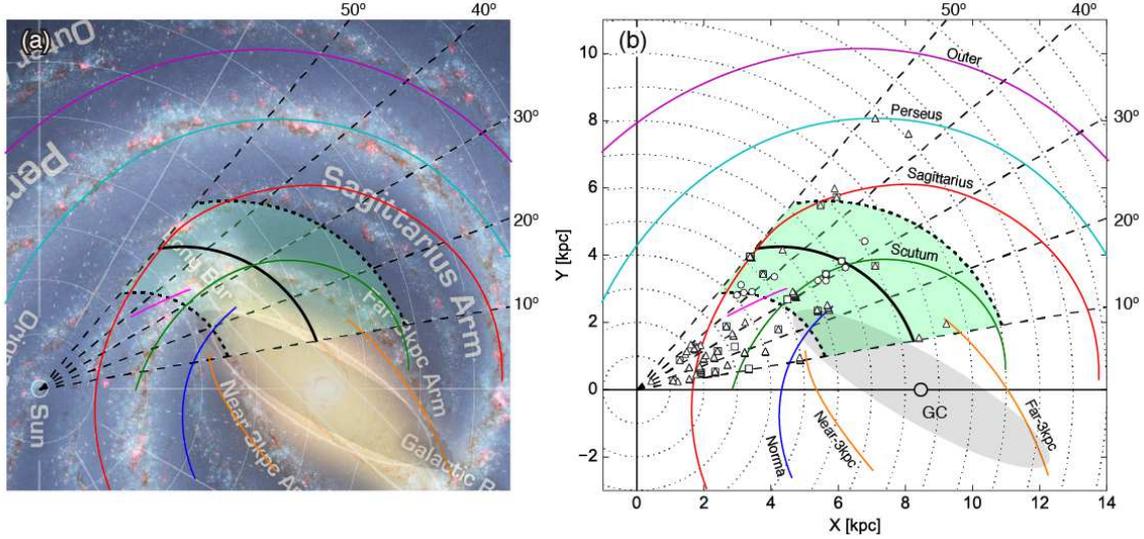}
 \end{center}
 \caption{(a) The target region of this study is plotted on an illustration of the face-on view of the MW (NASA/JPL-Caltech/R. Hurt (SSC/Caltech)). The thick black line indicates the tangential points of the Galactic rotation relative to LSR, whose distances from the Sun ($d_{\rm tan}$) are used to compute the physical parameters of CO emissions, while the two thick dashed black lines shows the lines of $\pm30\%$ of $d_{\rm tan}$. The solid green area indicates the target region of the present analyses. Thin black dashed lines show the coverage of the FUGIN observations in the first quadrant, which are plotted every $10^\circ$ from $l=10^\circ$ to $50^\circ$. Dotted lines show isodistance contours from the Sun. Colored lines denote loci of the Galactic arms constructed by \citet{rei2016}.
 (b) is the same as (a) but without the background image and with the plots of the sources whose distances are determined by trigonometry \citep{hou2014}, where the circles, squares, triangles depict the GMCs, maser sources, and H{\sc ii} regions, respectively. 
}\label{fig:faceon}
\end{figure}

In Figure\,\ref{fig:faceon}(a), the thick black line plotted on an illustration of the face-on view of the MW indicates the tangential points included in $l=10^\circ$--$50^\circ$.
The dashed black lines and solid green area indicate the area at distances within $\pm30\%$ of $d_{\rm tan}$, at which the $\mH$ of molecular gases were measured in this study.
This target area, measuring $\sim25.7$\,kpc$^{2}$, was defined to include the Galactic bar, Far-3\,kpc Arm, Norma Arm, Scutum Arm, and Sagittarius Arm as well as to satisfy the required conditions for measuring $\fDG$ by quantifying the $M_{\rm H_2}$ traced by $^{12}$CO, $^{13}$CO, and C$^{18}$O as discussed above.
At a $d_{\rm tan}$ $\sim5.5$--$8.4$\,kpc in this area, the $b$ coverage of the FUGIN data ($|b| \leq \pm1^\circ$) corresponds to $\sim192$--$293$\,pc (or 134--380\,pc including the $\pm30$\% error of $d_{\rm tan}$), and the spatial resolutions of the post-processed FUGIN data were calculated as $\sim1.1$--$1.6$\,pc (or 0.7--2.1\,pc including the $\pm30$\% error of $d_{\rm tan}$). 
 In the majority of the target area, the vertical coverages and spatial resolutions satisfied the required conditions for measuring $\fDG$; however in some parts at higher $l$ the vertical coverages are less than 200\,pc.
Thus, the vertical extent of the $^{12}$CO emission may not be fully covered in these parts. 

\begin{figure}
 \begin{center}
  \includegraphics[width=13cm]{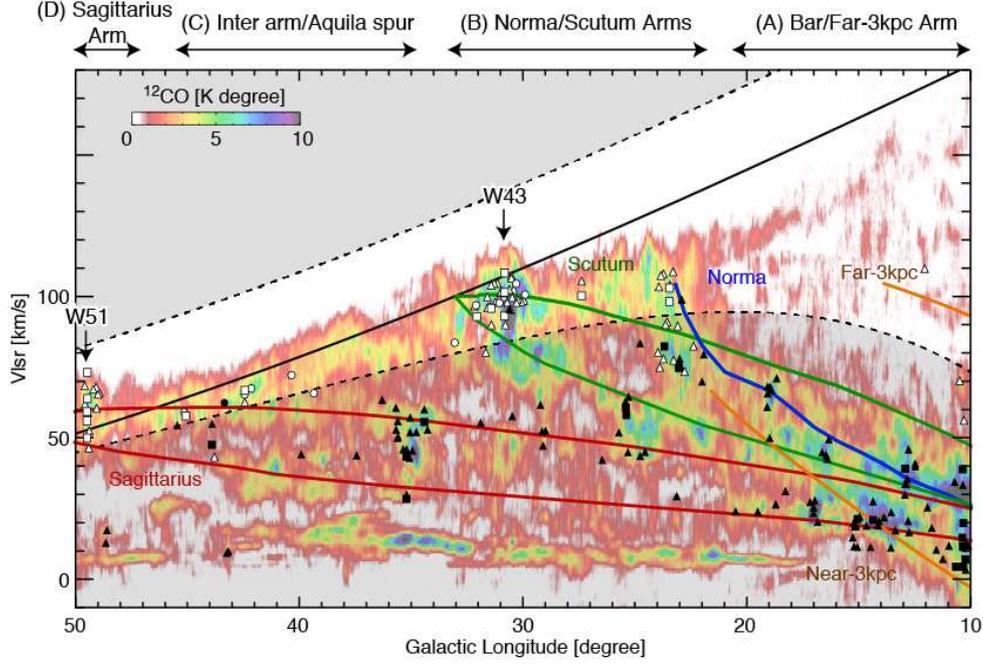}
 \end{center}
 \caption{The $l$-$v$ diagram of the FUGIN $^{12}$CO $J$=1--0 data. Integration range in $b$ is from $-1^\circ$ to $+1^\circ$.The thick black line shows a curve of the terminal velocities. The two dashed lines define the target velocity ranges of the present analyses; the one plotted below the curve of the terminal velocities indicates the velocities which correspond to $+30\%$ or $-30\%$ of $t_{\rm tan}$, while the other shows a $+30$\,km\,s$^{-1}$ margin from the terminal velocities. The masked area in this figure is not used for the $M_{\rm H_2}$ estimates in this study. The colored lines show the loci of the Galactic arms constructed by \citet{rei2016}. The four horizontal arrows show the rough extents of the four regions---Regions A--D---of this study. 
The circles, squares, and triangles show the GMCs, maser sources, and H{\sc ii} regions of the \citet{hou2014} samples, respectively. The open and closed symbols indicate the sources within and outside $d_{\rm tan} \pm 30$\% in Figure\,\ref{fig:faceon}, respectively.}
\label{fig:lv}
\end{figure}

Figure\,\ref{fig:lv} shows the $l$-$v$ diagram of the FUGIN $^{12}$CO $J$=1--0 data integrated over $\pm1^\circ$ in $b$.
The thick black line shows a curve of the $v_{\rm term}$.
The two dashed lines define the target velocity ranges of the present analyses; the dashed line plotted below the $v_{\rm term}$ indicates the velocities which correspond to $\pm30\%$ of $d_{\rm tan}$, while the other shows a $+30$\,km\,s$^{-1}$ margin from the $v_{\rm term}$, which is set to cover the CO features with $v_{\rm LSR}$ higher than $v_{\rm term}$.
In Figures\,\ref{fig:faceon}(b) and \ref{fig:lv} the sources whose distances were determined by trigonometry are plotted; this data was compiled by \citet{hou2014}. 
Many of the sources within $d_{\rm tan}\pm30$\% (open symbols) are distributed within the target velocity ranges indicated as the unmasked area in Figure\,\ref{fig:lv}, and the numbers of the false-positives (sources within $d_{\rm tan}\pm30$\% are distributed outside target velocities) and false-negatives (sources outside $d_{\rm tan}\pm30$\% (closed symbols) are located within the target velocities) are small.
The small number of the false-negatives supports the assumption of the flat rotation in the target area. 
Here it is notable that the distance-determined sources were not catalogued in the Galactic Bar region (Figure\,\ref{fig:faceon}), and the assumption of flat rotation may not apply in this region owing to the non-circular rotation of the Galactic Bar  (e.g., \cite{rag1999, sor2012}). 
This results in some fraction of the molecular gas distributed in $d_{\rm tan}\pm30$\% not having $v_{\rm LSR}$ within the target velocities. 
However, it is still probably rare for the molecular gas outside the bar region to contaminate the target velocities plotted in Figure\,\ref{fig:lv}.

In Figure\,\ref{fig:lv} the loci of the Galactic arms constructed by \citet{rei2016} are plotted in colored lines.
Two famous massive star forming regions---W51 and W43---are distributed around the tangential points of the Sagittarius Arm and Scutum Arm, respectively (e.g, \cite{car1998, meh1994, mot2014, sof2018}).
The 3\,kpc Arm is thought to be distributed at the same $R$ as the major axis of the Galactic Bar (Figure\,\ref{fig:faceon}(a)).
Although the location of the tangential point of the 3\,kpc Arm has not been confirmed, \citet{gre2011} proposed that it may be around $l\sim20^\circ$--$22^\circ$ in the first quadrant.
Given the distributions of these components in $l$, the target region of this study can be roughly classified into four subregions; 
(Region A) the Galactic bar and Far-3\,kpc arm ($l < 20^\circ$), (Region B) the Norma Arm and the Scutum Arm ($l\sim22^\circ$--$33^\circ$), (Region C) the inter-arm region between the Scutum Arm and the Sagittarius Arm ($l\sim35^\circ$--$45^\circ$), and (Region D) the Sagittarius Arm ($l>47^\circ$).
Note that these classifications remain ambiguous, as the distributions of the Galactic Bar and arms are not fully understood.

\begin{figure}
 \begin{center}
  \includegraphics[width=12cm]{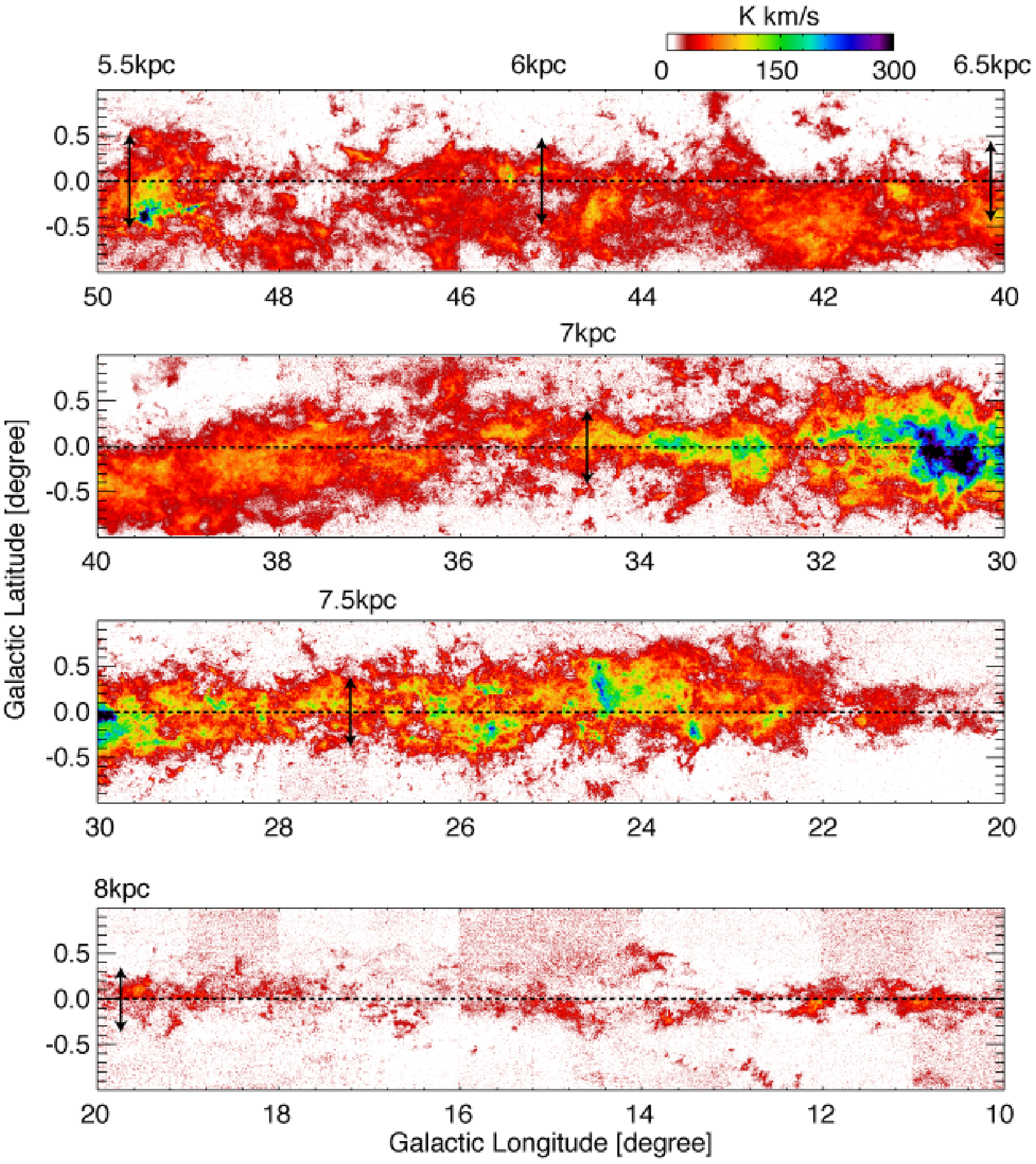}
 \end{center}
 \caption{Integrated intensity distributions of the FUGIN $^{12}$CO $J$=1--0 data for the velocity ranges plotted in Figure\,\ref{fig:lv} ($W({\rm ^{12}CO})$). The vertical arrows indicate the $\pm$50\,pc heights at $d_{\rm tan} = 5.5, 6, 6.5, 7, 7.5$, and $8$\,kpc. }\label{fig:lb12}
\end{figure}

\begin{figure}
 \begin{center}
  \includegraphics[width=12cm]{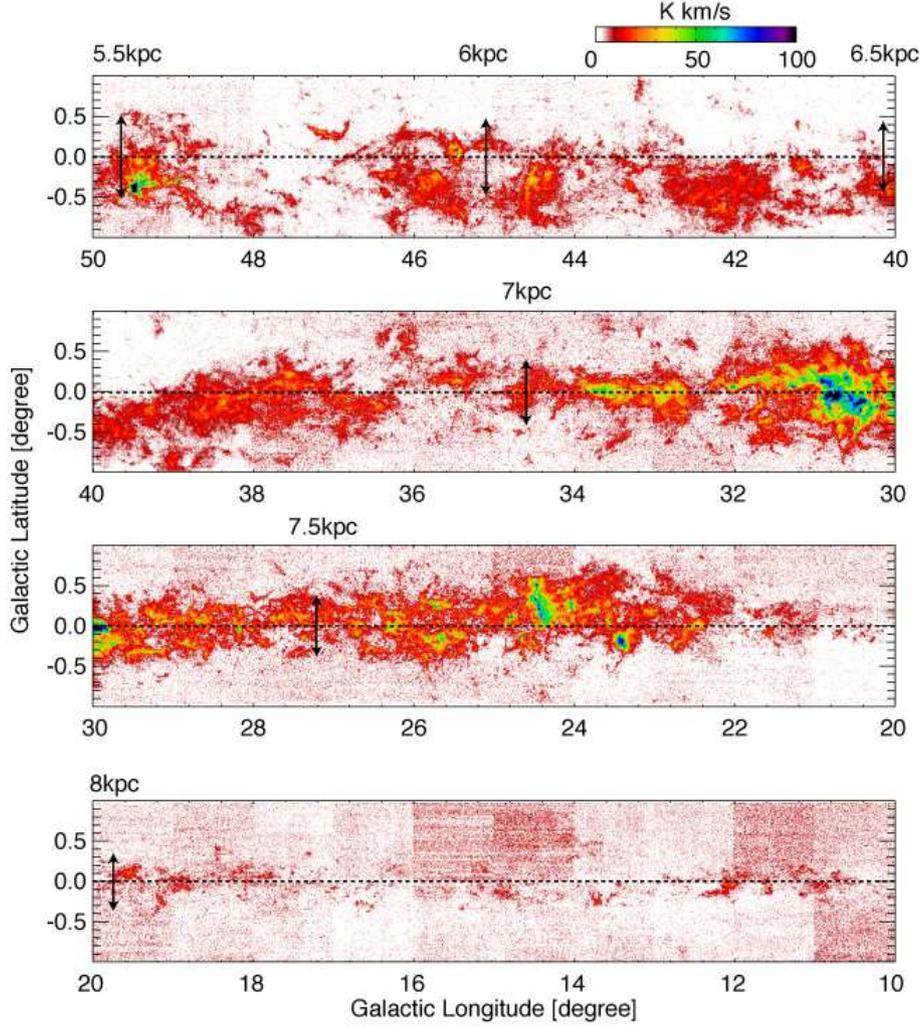}
 \end{center}
 \caption{Same as Figure\,\ref{fig:lb12}, but for $W({\rm ^{13}CO})$.}\label{fig:lb13}
\end{figure}

\begin{figure}
 \begin{center}
  \includegraphics[width=12cm]{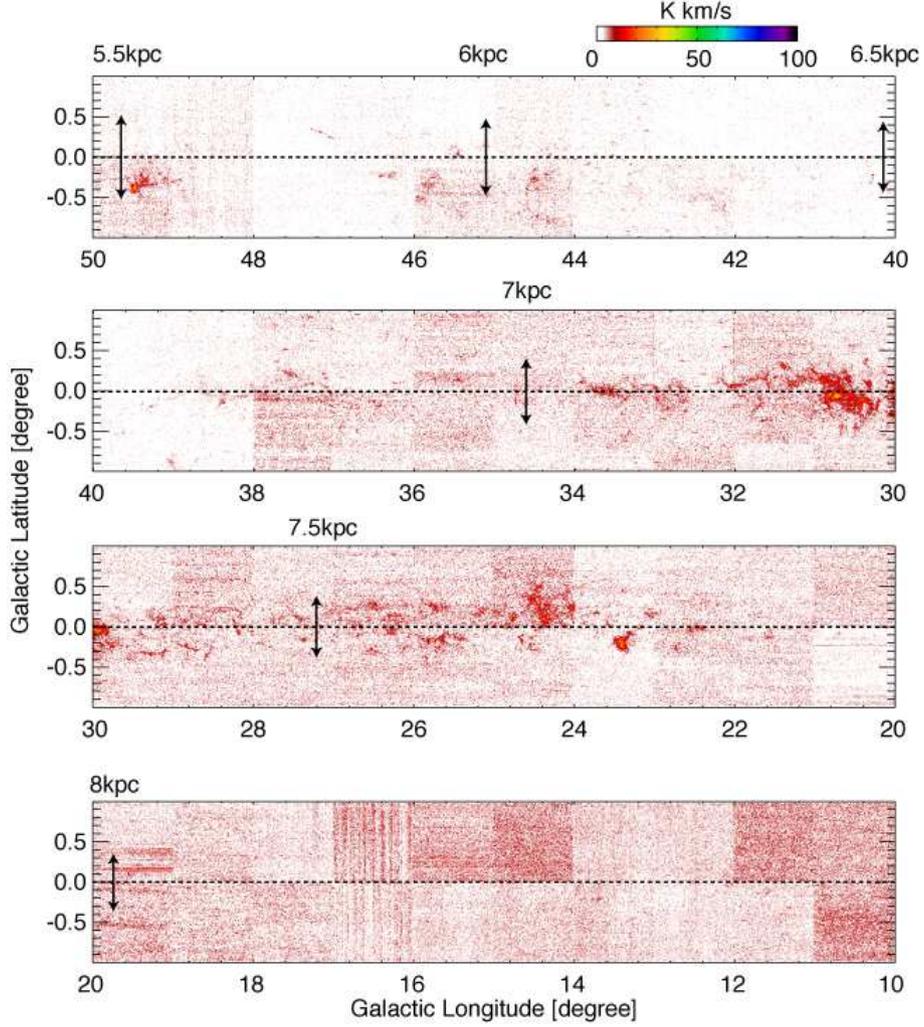}
 \end{center}
 \caption{Same as Figure\,\ref{fig:lb12}, but for $W({\rm C^{18}O})$.}\label{fig:lb18}
\end{figure}

\begin{figure}
 \begin{center}
  \includegraphics[width=10cm]{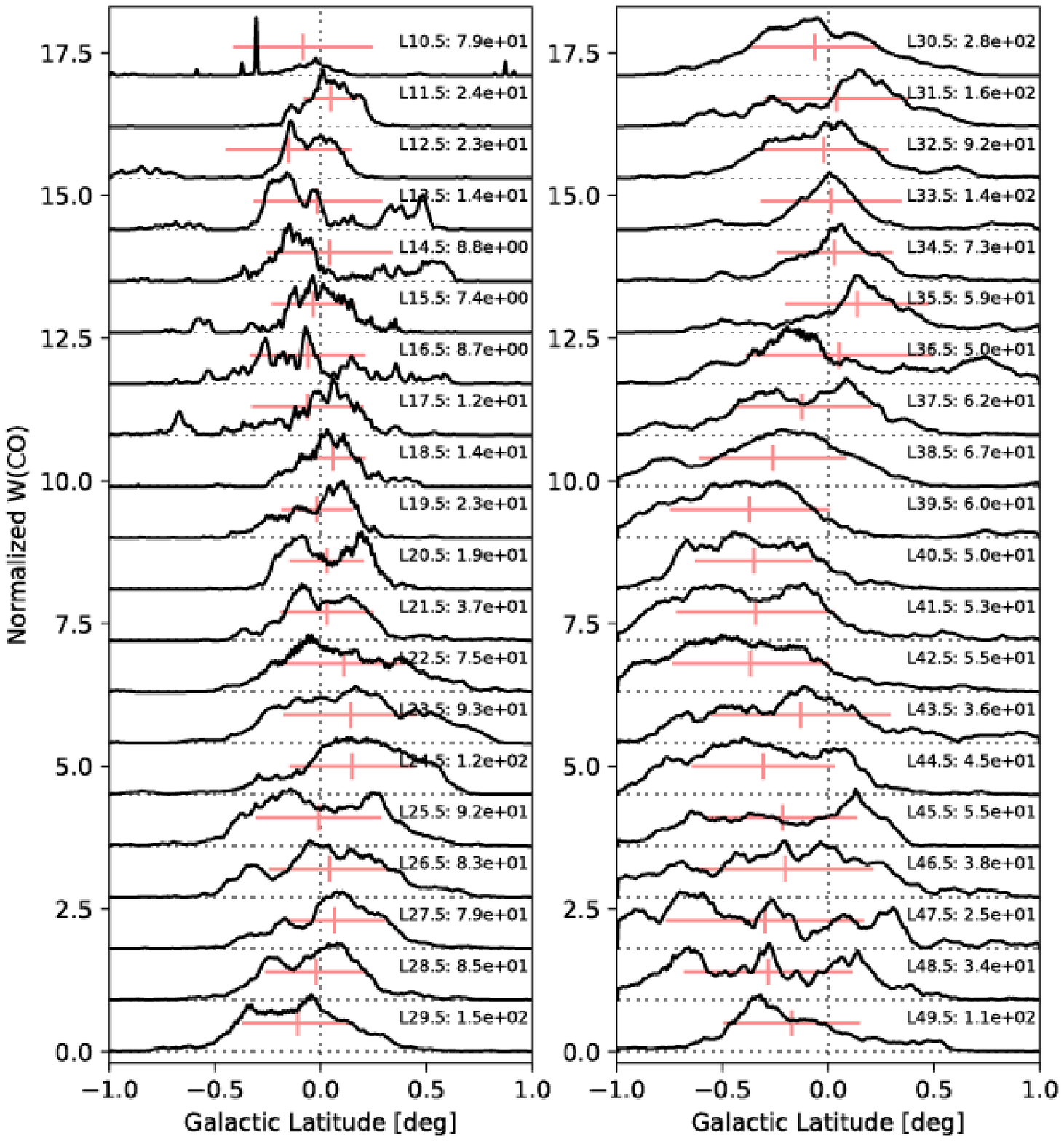}
 \end{center}
 \caption{The $W({\rm ^{12}CO})$ distributions along the Galactic latitude are plotted with a bin-size of $l=1^\circ$. The peak $W({\rm ^{12}CO})$ of each plot is normalized to 1, and the peak value is presented at the top-right of the plot in K\,km\,s$^{-1}$. The pink lines indicate the intensity-weighted average velocities and $\pm1\sigma$ velocity dispersions. }\label{fig:bdist}
\end{figure}

Figures\,\ref{fig:lb12}, \ref{fig:lb13}, and \ref{fig:lb18} show the $^{12}$CO, $^{13}$CO, and C$^{18}$O intensity distributions, respectively, integrated over the target velocity ranges of Figure\,\ref{fig:lv}. 
These distributions are denoted $W({\rm ^{12}CO}), W({\rm ^{13}CO}), W({\rm C^{18}O})$ hereon. 
It can be seen from Figures\,\ref{fig:lb12}--\ref{fig:lb18} that the vertical extents of the CO emission are well-covered within the $|b| < 1^\circ$ coverage of the FUGIN observations.
More detailed $b$ distributions of the $W({\rm ^{12}CO})$ are shown in Figure\,\ref{fig:bdist}, where the $W({\rm ^{12}CO})$ profiles along $b$ are plotted at every $1^\circ$ in $l$, with the intensity-weighted average velocities and $\pm1\sigma$ velocity dispersions plotted with pink lines.
This figure indicates that the vertical distributions of the $^{12}$CO emission are sufficiently covered to estimate the total $M_{\rm H_2}$ in all the regions except for $l\sim40\fdg5$--$48\fdg5$, where the $W({\rm ^{12}CO})$ distribution is shifted toward the negative direction in $b$ by $\sim0.2^\circ$--$0.3^\circ$, running off the edge at $b=-1^\circ$. 
As the $\pm1\sigma$ dispersions were covered even in these regions, it is expected that approximately 80--90\% of the total $M_{\rm H_2}$ is included within the present $b$ coverage.


\section{Methods}
\subsection{Mass Calculations}
The H$_2$ column density measured by $^{12}$CO, $N_{\rm H_2}({\rm ^{12}CO})$, was estimated using the $X$(CO), and this study adopted a uniform value of $2.0\times10^{20}$\,(K\,km\,s$^{-1}$)$^{-1}$\,cm$^{-2}$ (Equation\,\ref{eq0}).
The $^{13}$CO column density $N_{13}$ was estimated by assuming  Local Thermodynamic Equilibrium (LTE). 
The excitation temperature of the $^{13}$CO emission $T_{\rm ex, 13}$ was derived in each line-of-sight using the peak brightness temperature of the optically-thick $^{12}$CO emission,  $T_{\rm peak}(^{12}{\rm CO})$, assuming a common excitation temperature between $^{12}$CO and $^{13}$CO;
\begin{equation}
T_{\rm ex, 13}\ = \ \frac{5.53}{\ln\{ 1 + 5.53 / (T_{\rm peak}(^{12}{\rm CO}) + 0.819) \} }.\label{eq1}
\end{equation}
Then, the $^{13}$CO optical depth $\tau_{13}$ at each voxel can be computed by the following equation:
\begin{equation}
\tau_{13} \ = \ -\ln \left\{ 1 - \frac{T(^{13}{\rm CO})}{5.29 (J_{13}(T_{\rm ex, 13}) - 0.164) } \right\}, \label{eq2}
\end{equation}
where $J_{13}(T) \equiv 5.29 / [\exp (5.29/T) - 1]$ and $T(^{13}{\rm CO})$ is the brightness temperature of the $^{13}$CO emission in each voxel.
 $N_{13}$ was finally computed by integrating $\tau_{13}$ along the given velocity ranges as follows:
\begin{equation}
N_{13}\ = \ 2.42 \times 10^{14} \frac{T_{\rm ex, 13} + 0.87}{1- \exp(-5.29/T_{\rm ex, 13})} \int \tau_{13}\, dv. \label{eq3}
\end{equation}

The C$^{18}$O column density $N_{18}$ was also estimated assuming LTE.
As the optically thick $^{12}$CO traces different parts of the molecular clouds with the optically thin C$^{18}$O, it is difficult to assume a common excitation temperature between $^{12}$CO and C$^{18}$O.
Further, it is difficult to estimate the excitation temperature of C$^{18}$O, $T_{\rm ex, 18}$, from the $^{13}$CO spectra, because $^{13}$CO emission is not always optically thick toward the identified C$^{18}$O sources. Thus, in this study a uniform $T_{\rm ex, 18}$ of 10\,K was assumed as the typical temperature of dense gas (e.g., \cite{oni1996, sch2016}).
Then, the C$^{18}$O optical depth, $\tau_{18}$, and $N_{18}$ were then derived as follows:
\begin{equation}
\tau_{18} \ = \ -\ln \left\{ 1 - \frac{T({\rm C^{18}O})}{5.27 (J_{18}(T_{\rm ex, 18}) - 0.166) } \right\}, \label{eq4}
\end{equation}
\begin{equation}
N_{18} \ = \ 2.54 \times 10^{14} \frac{T_{\rm ex, 18} + 0.87}{1- \exp(-5.27/T_{\rm ex, 18})} \int \tau_{18} \, dv, \label{eq5}
\end{equation}
where $J_{18}(T) \equiv 5.27 / [\exp (5.27/T) - 1]$ and $T({\rm C^{18}O})$ is the brightness temperature of the C$^{18}$O emission.

The derived $N_{13}$ and $N_{18}$ in Equations \ref{eq3} and \ref{eq5} were then converted into the H$_2$ column densities, $N_{\rm H_2}({\rm ^{13}CO})$ and $N_{\rm H_2}({\rm C^{18}O})$, respectively, by adopting the abundance ratios of the CO isotopologues.
This study adopted the following relationships constructed by \citet{wil1994}:
\begin{equation}
[{\rm ^{12}C}]/[{\rm ^{13}C}] \ = \ (7.5\pm1.9) R + (7.6\pm12.9), \label{eq6}
\end{equation}
\begin{equation}
[{\rm ^{16}O}]/[{\rm ^{18}O}] \ = \ (58.8\pm11.8) R + (37.1\pm82.6). \label{eq7}
\end{equation}
The slope of the $[{\rm ^{12}C}]/[{\rm ^{13}C}]$ is consistent with the fits for the CO and CN data by \citet{mil2005}.
The measurements of the abundance ratios are not enough at $R < 4$\,kpc, while a $[{\rm ^{12}C}]/[{\rm ^{13}C}]$ of $20$ and a $[{\rm ^{16}O}]/[{\rm ^{18}O}]$ of $250$ were measured in the Galactic center at at $R<150$\,pc \citep{wil1994}.
Thus, the lower-limits for Equations\,\ref{eq6} and \ref{eq7} were set as 20 and 250 at $R < 4$\,kpc, respectively.
A [H$_2$]/[$^{12}$CO] ratio of $10^4$ was adopted (e.g., \cite{fre1982, leu1984}).
$N_{\rm H_2}({\rm ^{13}CO})$ and $N_{\rm H_2}({\rm C^{18}O})$ were finally derived based on the $[{\rm ^{12}C}]/[{\rm ^{13}C}]$ and $[{\rm ^{16}O}]/[{\rm ^{18}O}]$ calculated at the $R_{\rm tan}$ in each $l$.

The obtained $N_{\rm H_2}({\rm ^{12}CO})$, $N_{\rm H_2}({\rm ^{13}CO})$, and $N_{\rm H_2}({\rm C^{18}O})$ were then used to calculate the H$_2$ masses, $\mtwo$, $\mthree$, and $\meight$, respectively;
\begin{equation}
M_{\rm H_2} \ = \ 2.8 m_{\rm H} d_{\rm tan}^2 \Delta_{\rm grid}^2 N_{\rm H_2}, \label{eq8}
\end{equation}
where $m_{\rm H}$ is the mass of the hydrogen and $\Delta_{\rm grid}$ is the spatial grid-size of the CO data: 8.5$''$.

\subsection{Identifications of the CO sources}
The $\sigma$ of the FUGIN CO data displays a large variation for each region and for each CO line (Figure\,\ref{fig:rms}), and as presented in the histograms of the $\sigma$ in all the $1^\circ \times 2^\circ$ regions in Figures\,\ref{fig:sighis1}--\ref{fig:sighis5} in the appendix, in many regions the $\sigma$ distribution does not show symmetric profiles with respect to the average values, particularly in $^{12}$CO (Figures\,\ref{fig:sighis1}--\ref{fig:sighis5}); therefore it is difficult to estimate $\mH$ by simply summing all the pixel values included in the target velocity ranges. 
Hence, this study identified CO sources used to estimate $M_{\rm H_2}$ in the following two steps: (1) identify every local maximum by drawing contours in the data cube at a brightness temperature of $T_{\rm min}$, and (2) remove the identified structures with the voxel numbers less than $N_{\rm min}$ (the remaining structures are counted as CO sources).
The second step is useful for removing spurious structures; for instance, if $N_{\rm min}$ is sufficiently larger than the voxel number of the spurious structures \citep{ric2016}.
As the post-processed FUGIN CO data have a beamsize of 40$''$ with a grid size of 8.5$''$, approximately 20 pixels are included in the beam on the $l$-$b$ plane. Considering a narrow width (1--2 pixels) of the spurious features in velocity, a $N_{\rm min}$ value of $40$ was applied in the present identifications.
We found that this choice effectively removes many spurious structures that still remain after the post-processing (Section\,2).
Then $T_{\rm min}$ in the first step was determined with a fixed $N_{\rm min}$ of 40.

\begin{figure}
 \begin{center}
  \includegraphics[width=12cm]{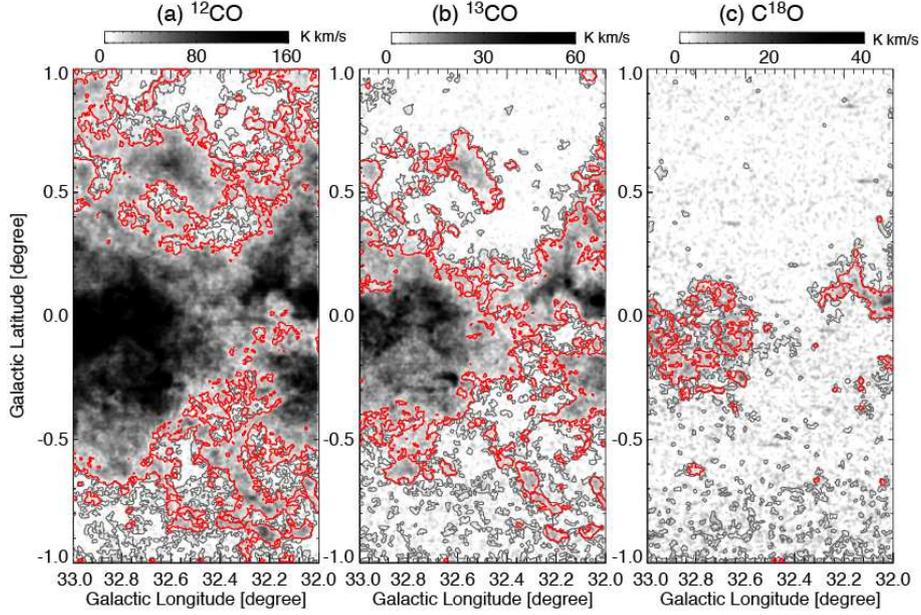}
 \end{center}
 \caption{Examples of CO source identifications at $l=32^\circ$--$33^\circ$ for (a) $^{12}$CO, (b) $^{13}$CO, and (c) C$^{18}$O data. The gray and red contours show the outlines of the identified structures at $T_{\rm min} = 3\sigma_{\rm med}$ and $5\sigma_{\rm med}$, respectively. The $1\sigma_{\rm med}$ of the $^{12}$CO, $^{13}$CO, and C$^{18}$O data were computed to be $\sim0.38$\,K, $\sim0.16$\,K, and $\sim0.16$\,K, respectively.}\label{fig:mask}
\end{figure}

Figure\,\ref{fig:mask} shows examples of the identifications of the $^{12}$CO, $^{13}$CO, and C$^{18}$O sources in the $1^\circ \times 2^\circ$ region at $l=32^\circ$--$33^\circ$.
Here, CO sources were identified using two $T_{\rm min}$ of $3\sigma_{\rm med}$ and $5\sigma_{\rm med}$, where $\sigma_{\rm med}$ is the median value of the $\sigma$ in this region.
As $\sigma$ increase at $b<-0\fdg5$ of this region (Figure\,\ref{fig:rms}), many tiny structures were identified at $b<-0\fdg5$ in all the three CO isotopologues at $T_{\rm min} = 3\sigma_{\rm med}$ (gray contours), which are misidentifications of the CO sources.
On the other hand, when $T_{\rm min} = 5\sigma_{\rm med}$ (red contours), these noise structures were removed, and the CO sources were properly identified.

\begin{figure}
 \begin{center}
  \includegraphics[width=6cm]{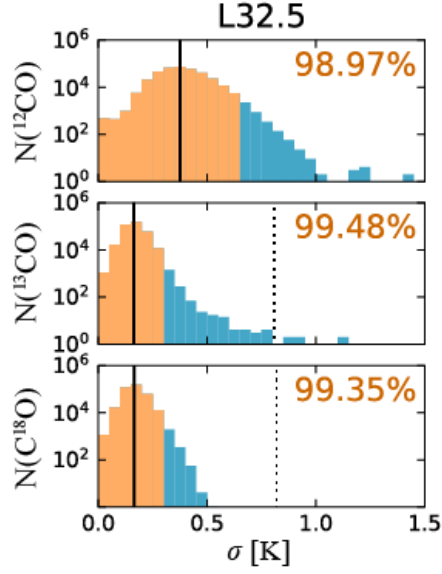}
 \end{center}
 \caption{
 Histograms of the $\sigma$ of the $^{12}$CO (top), $^{13}$CO (center), and C$^{18}$O data in the region at $l=32^\circ$--$33^\circ$ (Figure\,\ref{fig:mask}). Vertical solid and dotted lines respresent the $\sigma_{\rm med}$ and $5\sigma_{\rm med}$, respectively. The orange area indicates the data points with $3\sigma < 5\sigma_{\rm med}$, and the fraction of the data points in the orange area is shown at the top-right of the panel. The histograms of all the $1^\circ \times 2^\circ$ regions are presented in Figures\,\ref{fig:sighis1}--\ref{fig:sighis5} in the appendix.
 }\label{fig:sigmahist}
\end{figure}

Figure\,\ref{fig:sigmahist} shows the histograms of $\sigma$ of the $^{12}$CO, $^{13}$CO, and C$^{18}$O data in this region.
At the data points included in the orange areas the CO emissions were identified at $>3\sigma$ when $T_{\rm min} = 5\sigma_{\rm med}$, and these data points account for 98.97, 99.48, and 99.35\% of all the data points in this region for $^{12}$CO, $^{13}$CO, and C$^{18}$O, respectively, while these fractions decreased to 59.61, 79.63, and 78.70\% with $T_{\rm min} = 3\sigma_{\rm med}$, respectively.
CO sources could be identified at less than $3\sigma$ for many data points.
The $\sigma$ distributions in all the $1^\circ \times 2^\circ$ regions in Figures\,\ref{fig:sighis1}--\ref{fig:sighis5} in the appendix indicate that the $5\sigma_{\rm med}$ threshold can be used for the effective identification of CO sources.
Therefore, in this study $T_{\rm min} = 5\sigma_{\rm med}$ was uniformly adopted in a given region.
After applying this algorithm, the results of the identified structures were visually confirmed, and if the artificial structures due to the scanning effects were identified as the CO sources, these structures were removed by hand.

\subsection{Constructing the longitudinal distribution of $\mH$}
\begin{figure}
 \begin{center}
  \includegraphics[width=5.5cm]{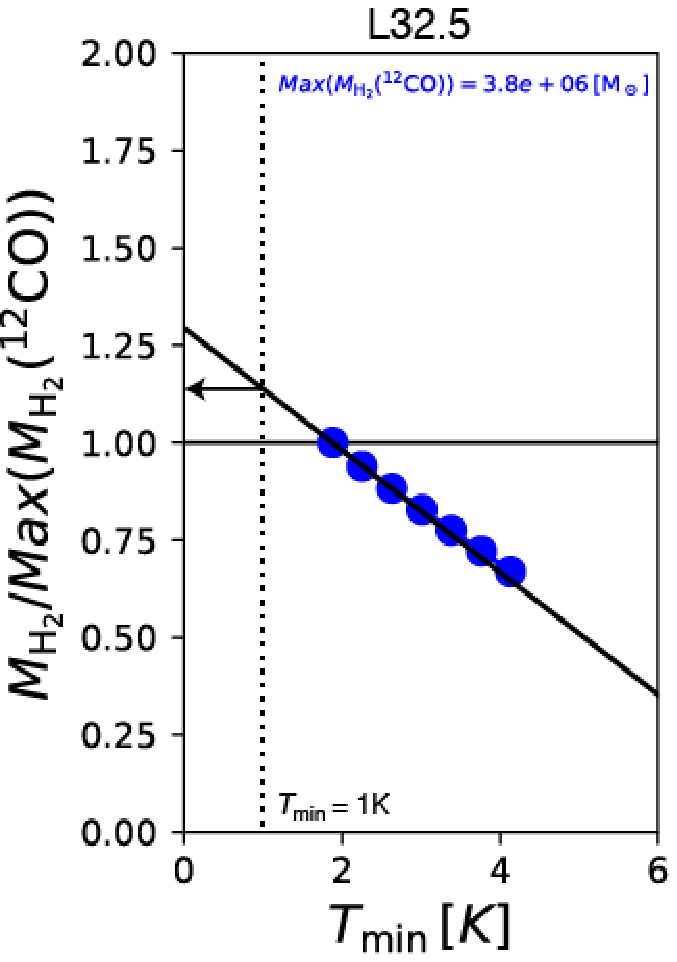}
 \end{center}
 \caption{An example of the extrapolation to derive the $\mtwo$ at $T_{\rm min} = 1$\,K at $l=32^\circ$--$33^\circ$. The blue circles indicates the normalized values of $\mtwo$ measured at $T_{\rm min} = 3, 4, 5, 6, 7, 8, 9, 10,$ and  $11 \times \sigma_{\rm med}$. The normalizations are generated by dividing the $\mtwo$ by $max(M_{\rm H_2}{\rm (^{12}CO)})$, which is the maximum $\mtwo$. The arrow indicates the resulting value of the extrapolation at $T_{\rm min} = 1$\,K. The plots of all the $1^\circ \times 2^\circ$ regions are presented in Figures\,\ref{fig:mplot1}--\ref{fig:mplot3} in the appendix.}\label{fig:expol}
\end{figure}

The target $40^\circ \times 2^\circ$ region was divided into forty tiles ($1^\circ \times 2^\circ$), and these tiles have different $5\sigma_{\rm med}$.
It is important to apply a uniform $T_{\rm min}$ to make fair comparisons between different tiles, and a $T_{\rm min}$ of $1$\,K was uniformly adopted in this study.
The $5\,\sigma_{\rm med}$ levels of the $^{13}$CO and C$^{18}$O data were lower than 1\,K in every tile, and the $\mthree$ and $\meight$ could be measured directly by applying $T_{\rm min} = 1$\,K.
However, the 5$\sigma_{\rm med}$ levels of the $^{12}$CO data exceed $1$\,K in all the tiles except for those with relatively low $\sigma$ at $l\sim38^\circ$--$43^\circ$.
A reasonable way of deriving the $\mtwo$ at $T_{\rm min} = 1$\,K in the data of tiles with high $\sigma$ is to apply an extrapolation technique.
The $\mtwo$ of the regions with $5\,\sigma_{\rm med} > 1$\,K at $T_{\rm min} = 5, 6, 7, 8, 9, 10$ and $11\times\sigma_{\rm med}$ were derived, and plots of the derived $\mtwo$ were made as functions of $T_{\rm min}$.
Figure\,\ref{fig:expol} shows an example at $l=32^\circ$--$33^\circ$,
where the vertical axis shows $\mtwo$ divided by $max(M_{\rm H_2}{\rm (^{12}CO)})$, which is the maximum $\mtwo$ measured at $T_{\rm min} = 3\,\sigma_{\rm med}$.
The plots were extrapolated by making linear-fits using the three data points at $T_{\rm min} = 5, 6,$ and $7\times\sigma_{\rm med}$ to estimate the $\mtwo$ at $T_{\rm min} = 1$\,K.
The results of the extrapolations in all the $1^\circ \times 2^\circ$ regions are presented in Figures\,\ref{fig:mplot1}--\ref{fig:mplot3} in the appendix.
The resulting $\mtwo$ was increased by a factor of $\sim1.1$--$2.0$ from the $max(M_{\rm H_2}{\rm (^{12}CO))}$.
If two or four data points were used for the fits instead of three, the obtained factors changed by $\pm1$--$7$\% depending on the temperature difference $\Delta T\,{\rm (K)}= 5\sigma_{\rm med}-T_{\rm min}$. 
It would be the best if $\mH$ at $T_{\rm min} = 0$\,K was calculated using this extrapolation technique, however, the errors in this case could become significantly large because of large $\Delta T$.
Therefore, $T_{\rm min} = 1\,$K was applied in this study to suppress the errors due to the extrapolations.

\subsection{Uncertainties}
\subsubsection{$\mtwo$}
The uncertainty of the derived $\mtwo$ was calculated from the uncertainties of $W{\rm (^{12}CO)}$, $X{\rm (CO)}$, and extrapolation.
\citet{ume2017} discussed that the FUGIN data has uncertainties of the observed brightness temperatures of $\pm10$--$20\%$ for $^{12}$CO and $\pm10\%$ for $^{13}$CO and C$^{18}$O \citep{ume2017}.
The uncertainty of the $X$(CO) was uniformly set as $\pm30$\% as reported by \citet{bol2013} for $R = 1$--$9$\,kpc. 
Note that the assumption of the uniform $X$(CO) possibly lead to overestimates of the derived $\mtwo$ at $R<\sim2$--$3$\,kpc by up to a factor of $\sim2$, as  it has been reported that $X$(CO) decreases in the Galactic center region at $R < 1$\,kpc, e.g., $0.24\times10^{20}$\,(K\,km\,s$^{-1}$)$^{-1}$\,cm$^{-2}$ \citep{oka1998} at $R<0.1$\,kpc and $0.7\times10^{20}$\,(K\,km\,s$^{-1}$)$^{-1}$\,cm$^{-2}$ at $R\sim0.7$\,kpc \citep{tor2010}.
As it is difficult to estimate the uncertainty of the extrapolations in Section\,4.3 including choice of the fitting function, a uniform error of $\pm20\%$ was assumed.
In addition, the derived $M_{\rm H_2}$ is affected by a distance error of $30\%$ as shown in Figure\,\ref{fig:faceon}, which can be canceled in taking ratios among $\mtwo$, $\mthree$, and $\meight$.

\subsubsection{$\mthree$ and $\meight$}
The uncertainties of the $N_{13}$ and $N_{18}$ were estimated from the uncertainties of the $T({\rm ^{13}CO})$ and $T({\rm C^{18}O})$ ($\pm10\%$; \cite{ume2017}) and the uncertainties of the $T_{\rm ex, 13}$ and $T_{\rm ex, 18}$, respectively (see Equations\,\ref{eq1}--\ref{eq5}).
Here, a uniform error of $\pm50$\% was assumed for $T_{\rm ex, 13}$ and $T_{\rm ex, 18}$, as it is not easy to evaluate these uncertainties in the large target area of this study, and then the $\pm30\%$ uncertainties of $N_{13}$ and $N_{18}$ were derived.
In converting $N_{13}$ and $N_{18}$ into $N_{\rm H_2}({\rm ^{13}CO})$ and $N_{\rm H_2}({\rm ^{13}CO})$, the uncertainties on the abundance ratios among the CO isotopologues were significant, which can be calculated as $\sim \pm40\%$ and $\sim \pm30\%$ for ${\rm [^{12}C]/[^{13}C]}$ and ${\rm [^{16}O]/[^{18}O]}$, respectively (Equations\,\ref{eq6} and \ref{eq7}).

In addition to these statistical uncertainties, a systematic error of $+20\%$ was considered for $N_{13}$ and $N_{18}$, as the LTE assumption may overestimate the true column densities due to the subthermal excitation of higher rotational transitions of CO \citep{har2004}.
Furthermore, in each $l$ the present analysis includes uncertainties of $R$ in Equations\,\ref{eq6} and \ref{eq7}, due to the $\pm30$\% error of $d_{\rm tan}$, which provide additional $+10$--$+30$\% uncertainties for the abundance ratios.

\section{Results}
\begin{figure}
 \begin{center}
  \includegraphics[width=12cm]{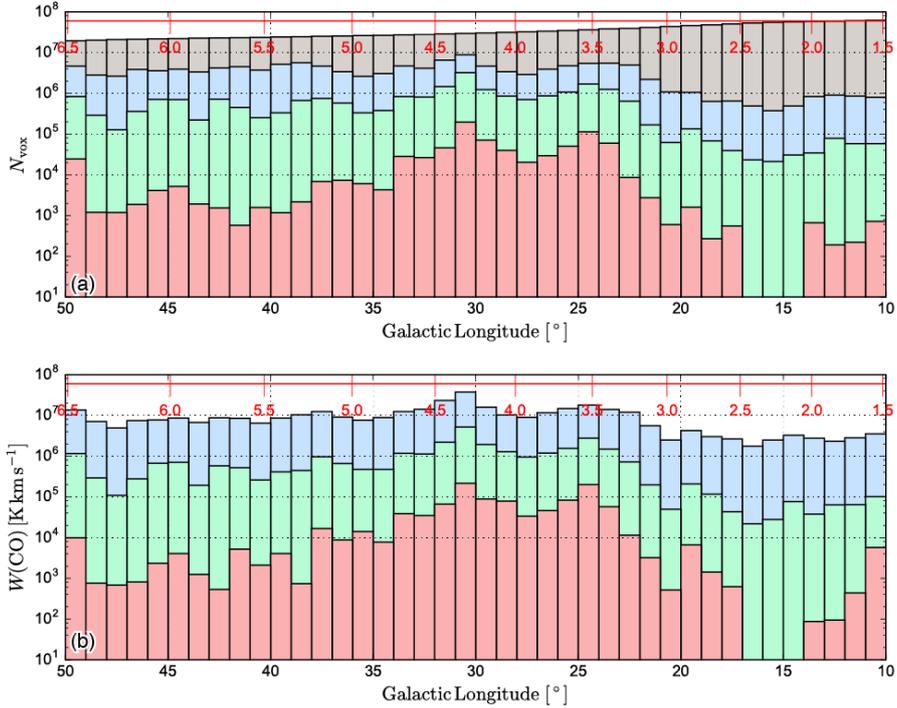}
 \end{center}
 \caption{The longitudinal distributions of (a)$N_{\rm vox}$ and (b)$W{\rm (CO)}$. The blue, green, and red bars show
the $^{12}$CO, $^{13}$CO, C$^{18}$O distributions, respectively. The horizontal red axis plotted in (a) indicates $R_{\rm tan}$. C$^{18}$O is not detected significantly in $l=14^\circ$--$17^\circ$. In (a) the gray bar indicates the total number of the voxels included in the target $l$ and $v$ ranges (Figure\,\ref{fig:lv}).}\label{fig:n+w}
\end{figure}

Figure\,\ref{fig:n+w} shows the longitudinal distributions of (a) the number of the voxels $N_{\rm vox}$ and (b) the $W({\rm CO})$ of the CO sources identified at $T_{\rm min} = 1$\,K.
The blue, green, and red bars indicate the derived values of $^{12}$CO, $^{13}$CO, and C$^{18}$O, respectively.
In addition, the gray bar in Figure\,\ref{fig:n+w} shows the total number of the voxels included in the target $l$ and $v$ ranges (Figure\,\ref{fig:lv}).
Here, the $N_{\rm vox}$ and $W({\rm CO})$ of the $^{12}$CO data (hereafter $N_{\rm vox}{\rm(^{12}CO)}$ and $W({\rm CO}){\rm(^{12}CO)}$) were derived by making extrapolations to $T_{\rm min} = 1$\,K, following the method described in Section\,4.3.
The $^{12}$CO and $^{13}$CO sources were detected in the entirety of the target region, while no C$^{18}$O sources were detected in the $l$ range of $14^\circ$--$17^\circ$ at $T_{\rm min} = 1$\,K.
The two peaks of $W({\rm CO})$ at $l=30^\circ$--$31^\circ$ and $49^\circ$--$50^\circ$ correspond to the regions that include W43 and W51, respectively.

\begin{figure}
 \begin{center}
  \includegraphics[width=12cm]{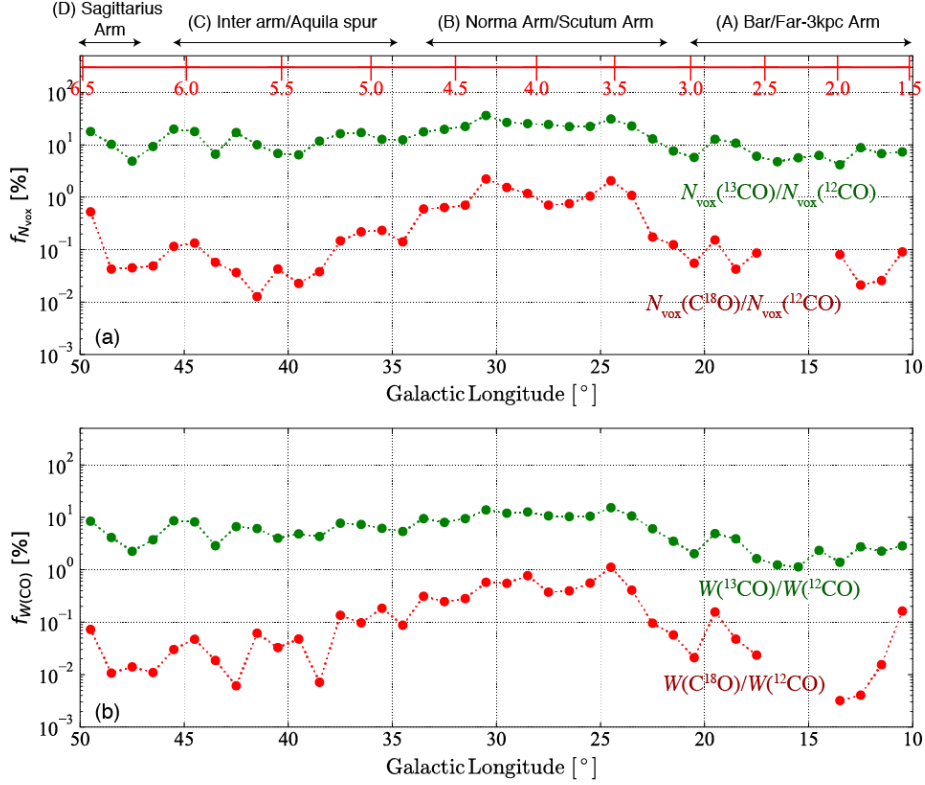}
 \end{center}
 \caption{The longitudinal distributions of (a)$f_{N_{\rm vox}}$ and (b)$f_{W{\rm (CO)}}$. The green circles indicate the ratios of $^{13}$CO to $^{12}$CO, while red circles show the ratios of C$^{18}$O to $^{12}$CO.
The horizontal red axis plotted in (a) indicates $R_{\rm tan}$.}\label{fig:fn+fw}
\end{figure}

Figure\,\ref{fig:fn+fw} shows the fractions of the $N_{\rm vox}$ and $W{\rm (CO)}$ plotted in Figures\,\ref{fig:n+w}(a) and (b) (hereafter $f(N_{\rm vox})$ and $f(W)$), respectively.
In Figure\,\ref{fig:fn+fw}(a) the $f(N_{\rm vox})$ of $^{13}$CO (green) and C$^{18}$O (red) to $^{12}$CO are presented ($f_{\rm ^{13}CO}({N_{\rm vox}})$ and $f_{\rm C^{18}O}({N_{\rm vox}})$, respectively), while in Figure\,\ref{fig:fn+fw}(b) the $f(W)$ of $^{13}$CO (green) and C$^{18}$O (red) to $^{12}$CO are plotted ($f_{\rm^{13}CO}(W)$ and $f_{\rm C^{18}O}(W)$, respectively).
The $f(N_{\rm vox})$ and $f(W)$ overall show similar distributions, with the $f(N_{\rm vox})$ being slightly larger (by a factor of $\sim$2--3) than the $f(W)$.
Figure\,\ref{fig:fn+fw} indicates that the $^{13}$CO, and particularly the C$^{18}$O emissions were detected at a small portion of the $^{12}$CO emitting regions.
The $f_{\rm ^{13}CO}({N_{\rm vox}})$ and $f_{\rm ^{13}CO}(W)$ range from $1\%$ to several 10\%, while the $f_{18}({N_{\rm vox}})$ and $f_{18}(W)$ range from $0.01$\,pc to $1\%$ with large variations.

\begin{figure}
 \begin{center}
  \includegraphics[width=13cm]{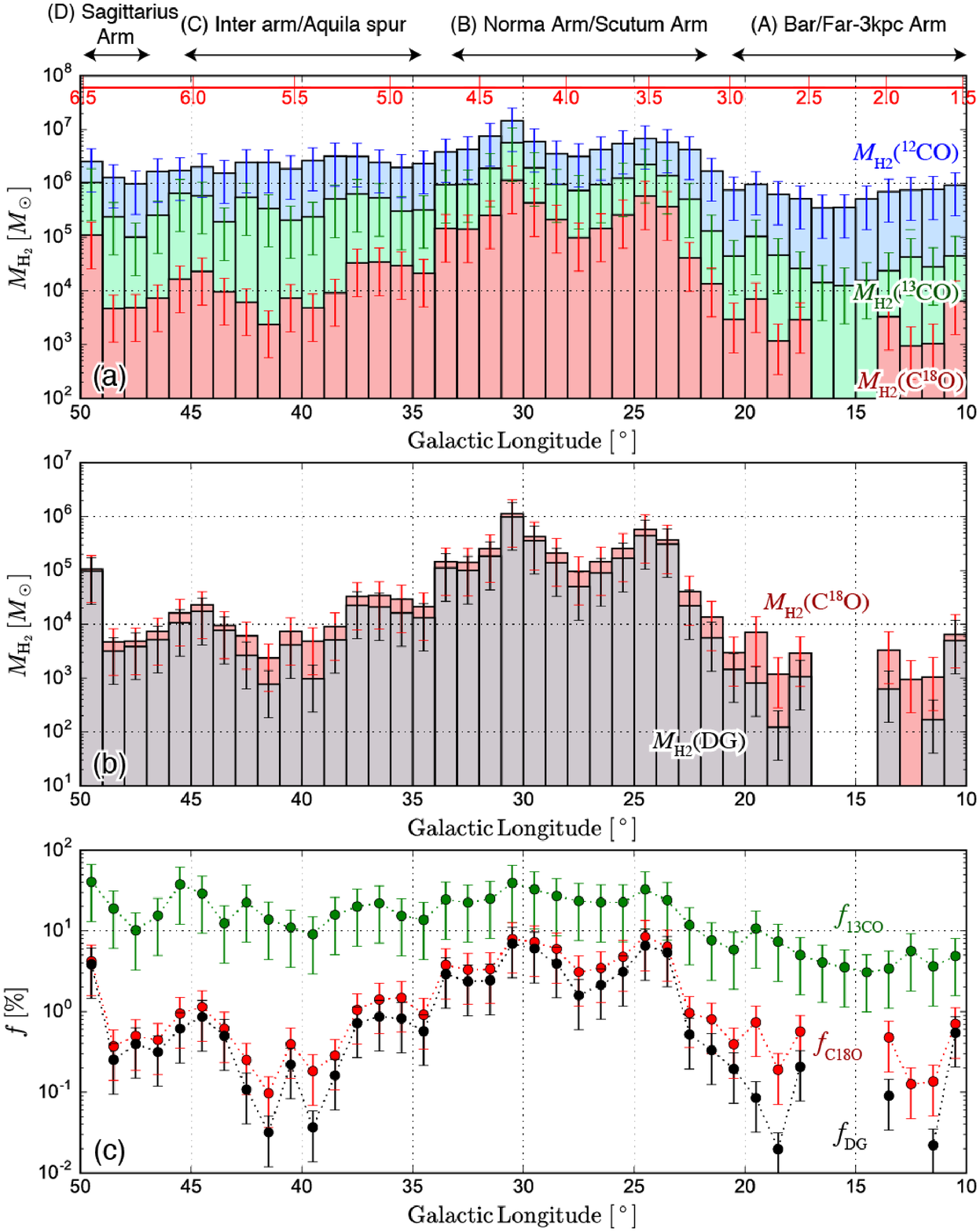}
 \end{center}
 \caption{(a) The longitudinal distribution of $\mH$ is shown in the same manner as Figure\,\ref{fig:n+w} but with error bars. (b) The distributions of $\meight$ and $\mDG$ are plotted with the red-bar and gray-bar, respectively. (c) Distributions of $\fthree$, $\feight$, and $\fDG$ are plotted in green, red, black, respectively.} \label{fig:m}
\end{figure}

Figure\,\ref{fig:m}(a) shows the longitudinal distribution of $M_{\rm H_2}$ in the same manner as Figure\,\ref{fig:n+w} but with error bars.
The $\mH$ of the dense gas $\mDG$ was calculated using the subregions of the identified C$^{18}$O sources at which $N_{\rm H_2}{\rm (C^{18}O)} \geq 7\times 10^{21}$\,cm$^{-2}$ (or $A_{\rm v} \geq 8$) and is plotted with gray-bars in Figure\,\ref{fig:m}(b).
The dense gas at which the filaments of 0.1\,pc width are dominant was not detected in $l=13^\circ$--$14^\circ$.

Figure\,\ref{fig:m}(c) shows the fractions of $\mthree$ (green), $\meight$ (red), and $\mDG$ (black) to $\mtwo$ in \% ($\fthree$, $\feight$, and $\fDG$, respectively).
The $\fthree$ in the inter-arm regions show slightly lower values than the Galactic arms, maintaining values of $\sim20$--$40\,\%$, while in Region A, which includes the Galactic Bar and Far-3kpc Arm, the $\fthree$ begins decreasing to $\sim4$--$10\%$.

On the other hand, the four regions show high variations in $\feight$ and, particularly in $\fDG$.
The fractions are relatively high in Regions B and D, ranging from $\sim2\%$ to $\sim8\%$, while the fractions are typically as low as $<1\,\%$ in Regions A and D, and some tiles have very low $\fDG$ of less than $0.1\%$.

Region B show two peaks in $\fDG$ at $l\sim30^\circ$ and $l\sim24^\circ$. The former corresponds to W43, while the latter includes a GMC associated with the infrared ring N35, which is an active star forming region \citep{tor2018a}.
These two star forming regions are probably located near the tangential points of Scutum Arm and Norma Arm, respectively, as seen in the $l$-$v$ diagram of Figure\,\ref{fig:lv}.
In addition, the $\feight$ and $\fDG$ increase in $l=49^\circ$--$50^\circ$ in Region D, where another active star forming region W51 is distributed around the tangential point of Sagittarius Arm.


The total $\mH$ of the three CO isotopologues in $l=10^\circ$--$50^\circ$ and their fractional masses are summarized in Table\,\ref{tab:1}; the derived $\mtwo$, $\mthree$, $\meight$, and $\mDG$ are $\sim10^{8.1}$\,$M_\odot$, $\sim10^{7.4}$\,$M_\odot$, $\sim10^{6.6}$\,$M_\odot$, and $\sim10^{6.5}$\,$M_\odot$, respectively. 
Given the surface area of  $25.7$\,kpc$^{2}$ of the target area of this study (Figure\,\ref{fig:faceon}), the corresponding surface mass densities are $\sim4.36$\,$M_\odot$\,pc$^{-2}$, $\sim1.02$\,$M_\odot$\,pc$^{-2}$, $\sim0.16$\,$M_\odot$\,pc$^{-2}$, and $\sim0.13$\,$M_\odot$\,pc$^{-2}$, respectively.
The averaged $\fthree$, $\feight$, and $\fDG$ are calculated as $23.7$\,\%, $3.7$\,\%, and $2.9$\,\%, respectively.

The total $\mH$ and fractional masses in the four regions, Regions A--D, are also summarized in Table\,\ref{tab:1}, where the borders of the two neighboring regions are removed.
In Regions A, B, C, and D, $\fthree$ has the average values of $5.5\%$, $29.9\%$, $17.2\%$, and $40.6\%$, while $\fDG$ has the average value of $0.1\%$, $4.8\%$, $0.4\%$, and $3.9\%$, respectively.
Region D has only one bin at $l=49^\circ$--$50^\circ$, resulting in a relatively large error on the averaged $\feight$ and $\fDG$.
Additional observations at $l \geq 50^\circ$ are needed to obtain more reliable representative values of the fractional masses in the Sagittarius Arm. 
In addition, as shown in Figure\,\ref{fig:bdist}, the vertical extents of the $^{12}$CO emission are not fully covered at $l\sim40^\circ$--$49^\circ$ in Region C, which may lead overestimating the obtained fractional masses by $\sim10$--$20\%$.

\begin{figure}
 \begin{center}
  \includegraphics[width=13cm]{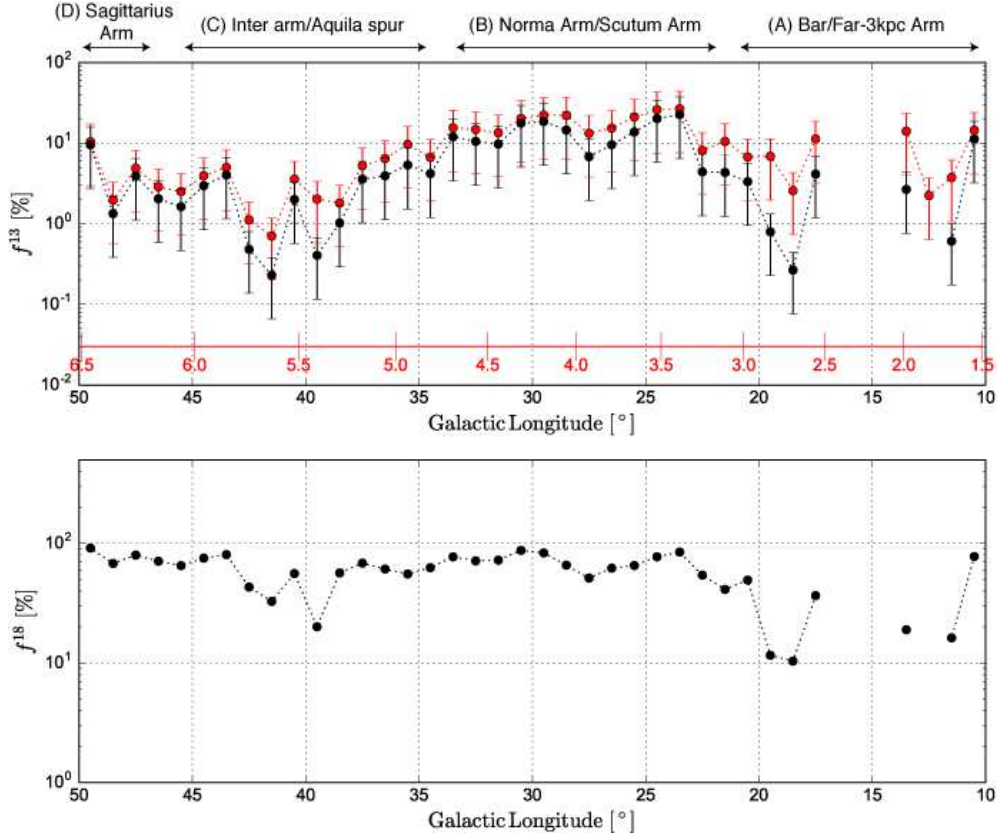}
 \end{center}
 \caption{The longitudinal distributions of $\ffeight$ (red) and $\ffDG$ (black) are plotted in (a), while the distribution of $f^{18}$ is shown in (b).}\label{fig:f13}
\end{figure}

Figure\,\ref{fig:f13} shows the fractions of $\meight$ (red) and $\mDG$ (black) to the $\mthree$ ($\ffeight$ and $\ffDG$, respectively) and the fractions of $\mDG$ to $\meight$ ($\fffDG$).
The averaged $\ffeight$, $\ffDG$, and $\fffDG$ in the entirety of $l=10^\circ$--$50^\circ$ and the four regions are summarized in Table\,\ref{tab:1}; 
The averaged $\ffeight$ and $\ffDG$ in $l=10^\circ$--$50^\circ$ are estimated as $15.7\%$ and $12.1\%$, respectively, while Regions A, B, C, and D have averaged $\ffeight$ of $6.4\%$, $20.5\%$, $3.9\%$, and $10.4\%$ and $\ffDG$ of $2.2\%$, $16.2\%$, $2.4\%$, and $9.5\%$, respectively.
The $\fffDG$ appears to be relatively stable compared with the $\fDG$ and $\ffDG$.
Although some regions in the Galactic Bar and inter-arm region show small $\fffDG$ of $\sim10\%$, while the other regions typically have high $\fffDG$ of $\sim70$--$90\%$.

Here, it is noted that, in this study the CO sources were detected at $T_{\rm min} = 1$\,K, and the derived fractions whose numerators are $\mDG$ (i.e., $\fDG$, $\ffDG$, and $\fffDG$) gave the upper-limits, as lower $T_{\rm min}$ would provide larger values for $\mtwo$, $\mthree$, and $\meight$, while the $\mDG$ was not changed.

\begin{table}
  \tbl{Total $M_{\rm H_2}$ and fractional masses}{
 \scriptsize
  \begin{tabular}{ccccccccccccccc}
  \hline
  \hline
     & & \multicolumn{4}{c}{$\log_{10} ( M_{\rm H_2} ) \ [M_\odot]$ } && \multicolumn{3}{c}{$f \ [\%]$} &&  \multicolumn{2}{c}{$f^{13} \ [\%]$}  &&  $f^{18} \ [\%]$\\
     \cline{3-6} \cline{8-10} \cline{12-13} \cline{15-15}
     Region & $l$ range & ${\rm ^{12}CO}$ & ${\rm ^{13}CO}$  & ${\rm C^{18}O}$ & dense gas && ${\rm ^{13}CO}$ & ${\rm C^{18}O}$ & dense gas &&  ${\rm C^{18}O}$ & dense gas && dense gas \\
     (1) & (2) & (3) & (4) & (5) & (6) &  & (7) & (8) & (9) && (10) & (11) && (12)\\
    \hline
    All & $10^\circ$--$50^\circ$ 	& $8.05^{+0.06}_{-0.07}$ 	&  $7.42^{+0.10}_{-0.11}$ 	& $6.62^{+0.12}_{-0.14}$ &  $6.51^{+0.12}_{-0.15}$ && $23.7^{+10.8}_{-10.9}$ &  $3.7^{+3.0}_{-3.0}$ & $2.9^{+2.6}_{-2.6}$ 	&& $15.7^{+8.5}_{-8.6}$ 	& $12.1^{+7.3}_{-7.3}$ 	&& $77.4^{+11.1}_{-11.1}$\\
    A   & $10^\circ$--$20^\circ$ 	& $6.81^{+0.09}_{-0.12}$ 	&  $5.55^{+0.15}_{-0.16}$ & $4.36^{+0.19}_{-0.19}$ &  $3.89^{+0.28}_{-0.31}$ && $5.5^{+2.8}_{-2.8}$ 	&  $0.4^{+0.3}_{-0.3}$ & $0.1^{+0.2}_{-0.2}$ 	&& $6.4^{+5.1}_{-5.2}$ 	& $2.2^{+3.5}_{-3.5}$ 	&& $35.6^{+28.3}_{-28.3}$\\
    B   & $23^\circ$--$32^\circ$ 	& $7.75^{+0.10}_{-0.14}$ 	&  $7.23^{+0.14}_{-0.18}$ & $6.54^{+0.13}_{-0.17}$ &  $6.44^{+0.14}_{-0.18}$ && $29.9^{+10.1}_{-10.5}$ &  $6.1^{+2.3}_{-2.5}$ & $4.8^{+2.3}_{-2.3}$ 	&& $20.5^{+7.0}_{-7.4}$ 	& $16.2^{+6.2}_{-6.5}$	&& $78.8^{+9.5}_{-9.5}$\\
    C   & $35^\circ$--$45^\circ$ 	& $7.38^{+0.09}_{-0.12}$ 	&  $6.61^{+0.11}_{-0.14}$ & $5.20^{+0.12}_{-0.16}$ &  $4.99^{+0.12}_{-0.16}$ && $17.2^{+6.7}_{-6.9}$ 	&  $0.7^{+0.5}_{-0.5}$ & $0.4^{+0.4}_{-0.4}$  	&& $3.9^{+2.7}_{-2.7}$ 	& $2.4^{+1.7}_{-1.8}$  	&& $61.9^{+12.1}_{-12.1}$\\
    D   & $49^\circ$--$50^\circ$ 	& $6.41^{+0.24}_{-0.57}$ 	&  $6.02^{+0.26}_{-0.71}$ & $5.03^{+0.25}_{-0.62}$ &  $4.99^{+0.25}_{-0.62}$ && $40.6^{+26.3}_{-27.5}$&  $4.2^{+2.3}_{-2.6}$ & $3.9^{+2.3}_{-2.4}$ 	&& $10.4^{+6.8}_{-7.4}$ 	& $9.5^{+6.2}_{-6.8}$ 	&& $91.5$\\
    \hline
  \end{tabular} }\label{tab:1}
   \begin{tabnote}
   (1) Region name. (2) Galactic longitude range used for the calculations of the average mass and fractional mass. (3--6) Logarithm of $M_{\rm H_2}$. (7--9) Fractional mass with $\mtwo$ as dinominator. (10, 11) Fractional mass with $\mthree$ as dinominator. (12) Fractional mass with $\mtwo$ as dinominator and $\mDG$ as numerator. 
   \end{tabnote}
\end{table}

\section{Discussion}
The $\mDG$ derived in this study is expected to be nearly consistent with the masses of the supercritical filaments \citep{and2014}, whose SFE has been found to be as quasi-universal throughout the galaxies (e.g., \cite{lad2010,shi2017}).
The analysis of the FUGIN CO data provides an average $\fDG$ value of $\sim2.9\%$ over $\sim5$\,kpc in the Galactic plane in the first quadrant (Table\,\ref{tab:1}).
This figure is consistent with the gap between the gas consumption time scale of $\sim1$--$2$\,Gyr (given by the KS-law) and the dense gas consumption timescale of $\sim20$\,Myr.
This suggests that the formation of dense gas in molecular clouds is the primary cause of inefficient star formation in galaxies, 
and which is consistent with the discussion by \citet{lad2010} that $\Sigma_{\rm SFR} \propto f_{\rm DG}\Sigma_{\rm H_2}$ is the fundamental relationship governing star formation.

On the other hand, the analyses of the FUGIN data revealed that there are huge variations of $\fDG$ depending on the structures of the MW disk.
In the regions including the Galactic arms (i.e., Regions B and D), $\fDG$ is as high as $\sim4$--$5\%$, while in the Galactic Bar and inter-arm regions (i.e., Regions A and C) it becomes quite small at $\sim0.1$--$0.4\%$.
As $\fDG$ is an indicator of the dense gas formation speed in the steady-state and the SFE in dense gas is likely quasi-universal, these large gaps in $\fDG$ may result in SFR differences between these regions.
Indeed, studies of the extra-galaxies indicate a systematic offset of $\sim50\%$ in the SFR among the arms, inter-arm, and bar (e.g., \cite{big2008, mom2010}). 
This is qualitatively consistent with our results, although the $50\%$ difference is smaller than the one order of magnitude difference found in this study.
Therefore, It is important to directly quantify the SFRs in the present target regions in the MW based on the infrared/radio observations.


The variations of the fractional masses may be attributed to the differences of the formation and destruction processes of the dense gas in the molecular clouds.
Although there are many theoretical studies on formation process of supercritical filaments (e.g., \cite{inu2015,fed2010,hen2008}), the fractional mass of the dense gas to total molecular gas has not yet been quantified.
The analyses first resolved the fractional masses over 5-kpc in the Galactic plane, which will encourage theoretical developments to understand the detailed process of dense gas formation in molecular clouds in the Galactic plane. 

A reasonable process for dense gas formation is compression by shock-wave.
\citet{inu2015} proposed a scenario of star formation for scales of $\sim100$\,pc, in which the multiple-compression of gas powered by the feedbacks of the massive stars regulates star formation in galaxies.
\citet{kob2017a} and \citet{kob2017b} constructed a semi-analytical model of GMC formation including the multiple-compressions driven by a network of expanding shells due to H{\sc ii} regions and supernova remnants, which resulted in finding slopes of the GMC mass functions similar as observed in spiral galaxies.

The roles of the galactic-scale gas motion have been discussed previously in the studies of extra-galaxies.
According to the spiral density wave theory \citep{shu2016}, the gas in the arms is affected by the strong compression caused by galactic shocks or cloud-cloud collisions. 
This mechanism can be expected in the other models, such as the non-steady spiral arm model \citep{wad2011, bab2013, dob2014}.
It has been suggested that the decrease in gas density observed in the bar regions of extra-galaxies can be attributed to the gravitationally unbound conditions of molecular clouds (e.g., \cite{sor2012, mei2013}): these conditions may be caused by the shear motion and/or cloud-cloud collisions (e.g., \cite{fuj2014}). 
\citet{yaj2018} discussed that the large velocity dispersion at $> 100$\,km\,s$^{-1}$ in the galactic bars may disperse GMCs. 
The decrease in the $\fthree$, $\feight$, and $\fDG$ in Region A may possibly be interpreted by these mechanisms.

The analyses of the FUGIN data have found no significant differences between the fractional masses of Regions A and C, except for $\fthree$, which showed the average values of $\sim6\%$ and $\sim17\%$ in Regions A and C, respectively (Figure\,\ref{fig:m}(c) and Table\,\ref{tab:1}).
This may suggest that formation of the relatively dense gas traced in $^{13}$CO is more efficient in the inter-arm regions rather than in the Galactic Bar, although there are no significant differences in $\feight$ and $\fDG$ between these two regions.
Observations of the extra-galaxies suggested that moderate shear motion in the arms may allow GMCs to stream into the inter-arm regions,  while GMCs hardly survive in the Galactic Bar (e.g, \cite{kod2009, miy2014}).
This may lead to higher $\fthree$ in the inter-arm region compared to the Galactic Bar.


To reach to a comprehensive understanding of the dense gas and star formation in the MW, it is important to perform additional analyses of the FUGIN CO dataset to identify and quantify the various structures of molecular gas in various spatial scales from 1\,pc to kpc, which will also allow us to make direct comparisons with the future large-scale observations of extra-galaxies with pc-scale resolutions.

\section{Summary}
The conclusions of the present study are summarized as follows.
\begin{enumerate}
\item The CO $J$=1--0 data, which was obtained as a part of the FUGIN project using the Nobeyama 45-m telescope, was analyzed to construct the longitudinal distributions of the $\mH$ traced by the $^{12}$CO, $^{13}$CO, and C$^{18}$O emissions with a bin-size of $l = 1^\circ$. 
\item $\mH$ was measured in the region within $d_{\rm tan} \pm 30\%$ by choosing the corresponding $v_{\rm LSR}$ ranges in the $l$-$v$ diagram. The target region included the Galactic Bar, Far-3kpc Arm, Norma Arm, Scutum Arm, Sagittarius Arm, and inter-arm regions. 
\item The $\mtwo$ of these regions were measured assuming the constant $X$(CO), and $\mthree$ and $\meight$ were estimated assuming LTE.  $\mDG$ was measured using the subregions of the C$^{18}$O sources at which $A_{\rm v} > 8$ mag.
\item The derived $\mDG$ and $\mtwo$ were then used to calculate $\fDG$, and the derived $\fDG$ showed large variations depending on the structures of the MW disk; the regions including the Galactic arms have high $\fDG$ of $\sim4$--$5\%$, while the $\fDG$ of the Galactic bar and inter-arm regions are small at $\sim0.1$--$0.4\%$. 
The averaged $\fDG$ over the entirety of the target region ($\sim5$\,kpc) is $\sim2.9\%$. This figure is consistent with the gap between the gas consumption timescale observed in the KS-law ($\sim1$--$2$\,Gyer) and dense gas consumption timescale ($\sim20$\,Myr), indicating that the formation of dense gas is the primary bottleneck of star formation in the MW.
\item Other mass ratios such as $\ffDG$ and $\fffDG$ were also measured; it was demonstrated that every mass ratio tends to increase in the arm regions as opposed to in the inter-arm and bar regions. 
Only $\fthree$ showed moderate differences between the arms and inter-arms, while still showing significantly small values in the bar region.
\item The analyses first resolved the $\fDG$ and other mass ratios over $\sim$5\,kpc in the Galactic plane, which provided crucial information on dense gas and star formation in the MW. It is expected that these results will encourage the future theoretical and observational studies.
\end{enumerate}

\begin{ack}
This work was financially supported by Grants-in-Aid for Scientific Research (KAKENHI) of the Japanese society for the Promotion of Science (JSPS; grant numbers 15H05694, 15K17607, 24224005, 26247026, and 23540277). 
Data analysis of the CO emissions was in part carried out on the open use data analysis computer system at the Astronomy Data Center, ADC, of the National Astronomical Observatory of Japan.
\end{ack}


\appendix

\section{Noise distributions of the FUGIN CO data}
In Figures\,\ref{fig:sighis1}--\ref{fig:sighis5} the histograms of the post-processed CO data are presented for all the $1^\circ \times 2^\circ$ regions in the same manner as in Figure\,\ref{fig:sigmahist}.

\section{Mass estimates from the $^{12}$CO data by extrapolations}
An extrapolation technique was adopted in this study to estimate the $\mtwo$ at $T_{\rm min} = 1$\,K (see Section\,4.3).
In Figures\,\ref{fig:mplot1}--\ref{fig:mplot3} the results of the extrapolations in all the $1^\circ \times 2^\circ$ regions analyzed in this study are presented in the same manner as in Figure\,\ref{fig:expol}.

\begin{figure}
 \begin{center}
  \includegraphics[width=13cm]{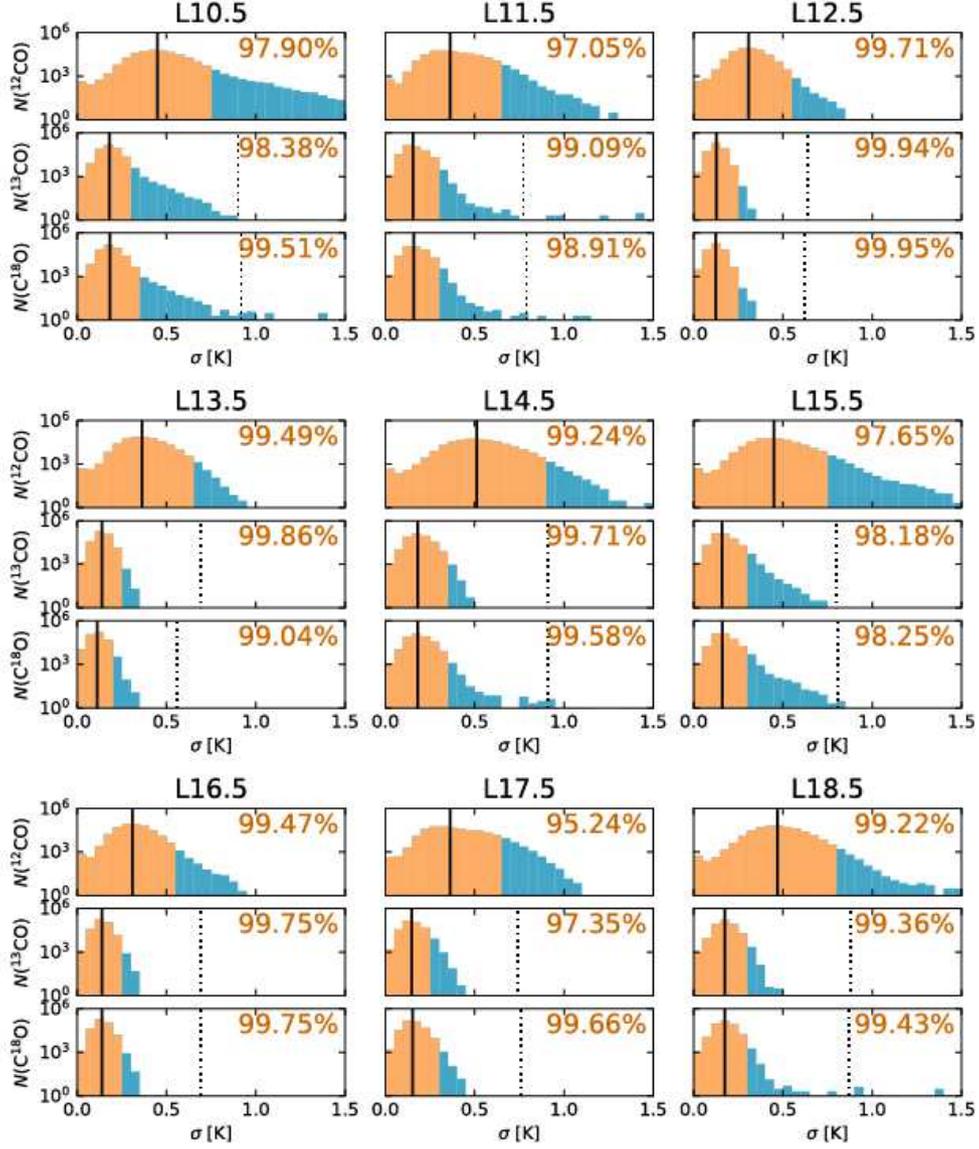}
 \end{center}
 \caption{Histograms of the $\sigma$ of the $^{12}$CO (top), $^{13}$CO (center), and C$^{18}$O data in all the $1^\circ \times 2^\circ$ regions analyzed in this study. Vertical solid and dotted lines indicate the $\sigma_{\rm med}$ and $5\sigma_{\rm med}$, respectively. The orange area indicates the data points with $3\sigma < 5\sigma_{\rm med}$, and the fraction of the data points in the orange area is shown at the top-right of the panel.}\label{fig:sighis1}
\end{figure}

\begin{figure}
 \begin{center}
  \includegraphics[width=13cm]{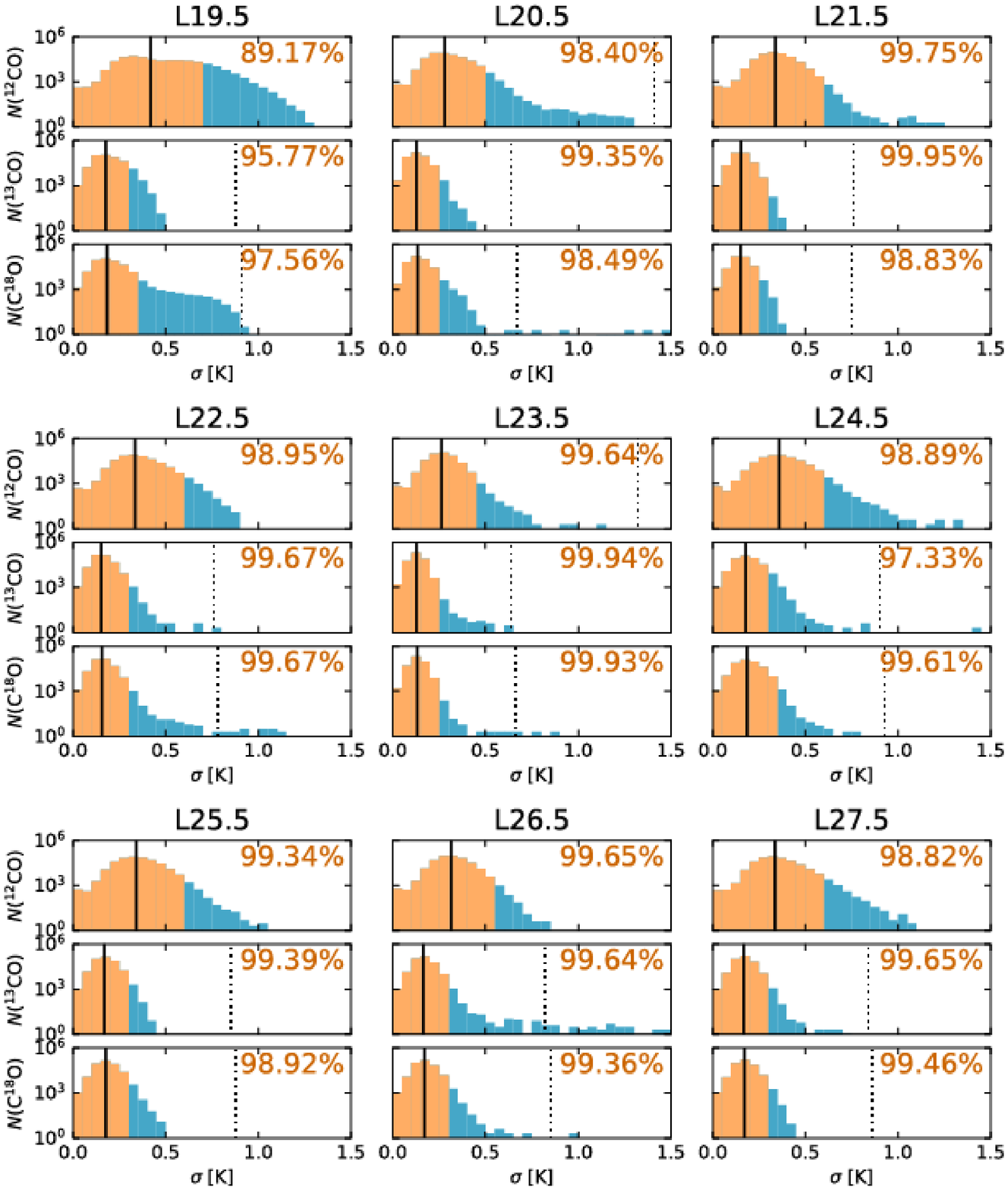}
 \end{center}
 \caption{Continued.}\label{fig:sighis2}
\end{figure}

\begin{figure}
 \begin{center}
  \includegraphics[width=13cm]{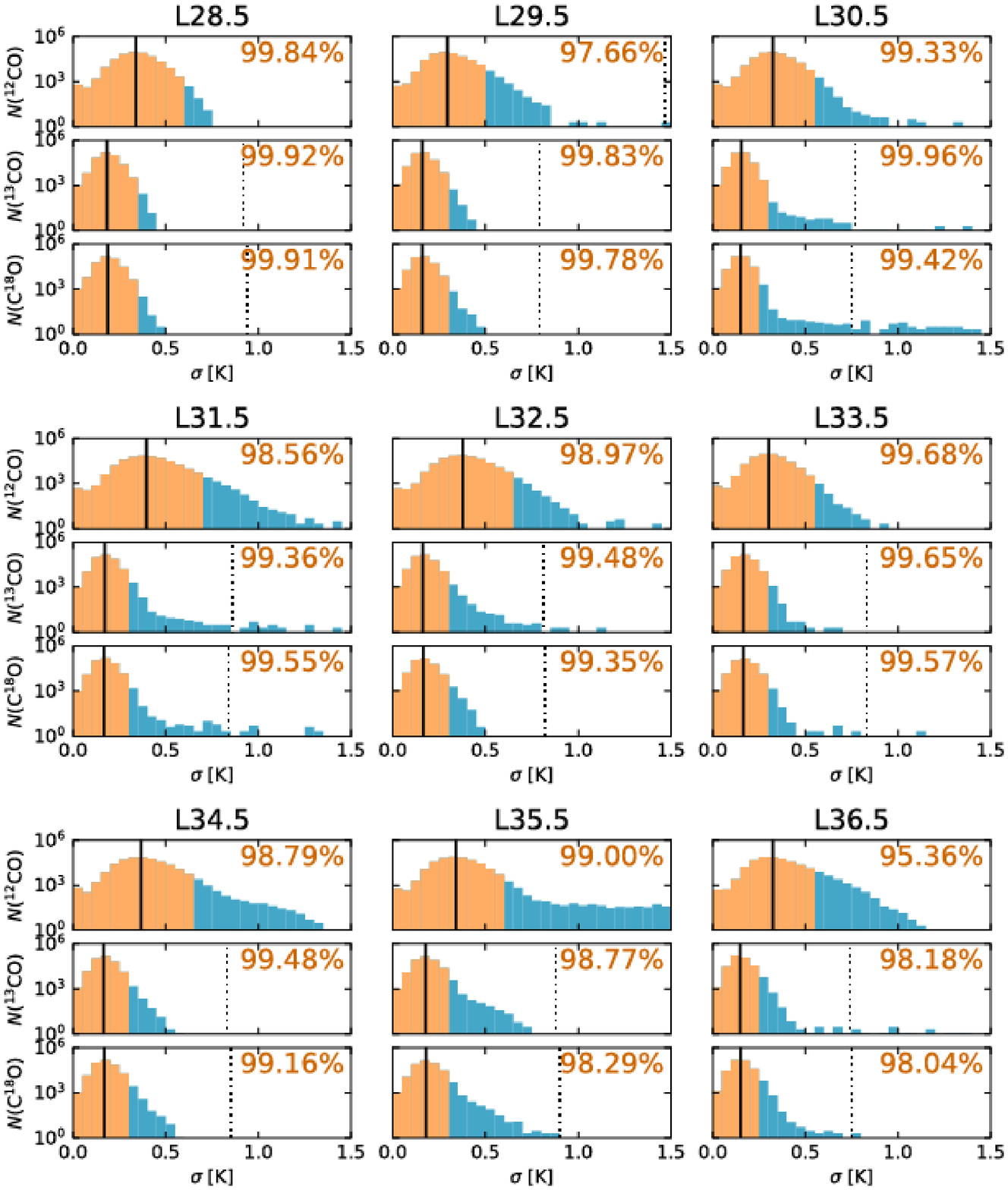}
 \end{center}
 \caption{Continued.}\label{fig:sighis3}
\end{figure}

\begin{figure}
 \begin{center}
  \includegraphics[width=13cm]{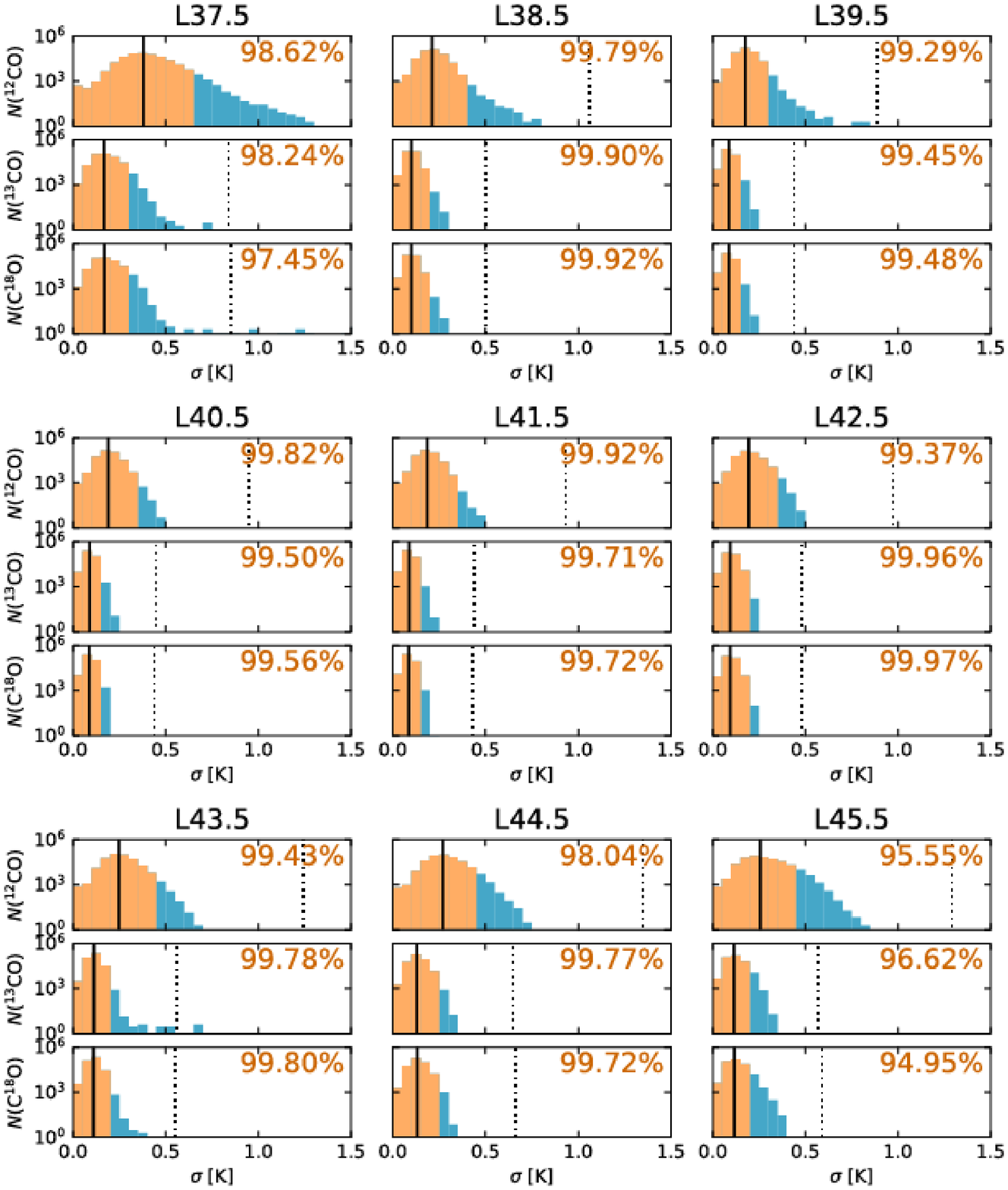}
 \end{center}
 \caption{Continued.}\label{fig:sighis4}
\end{figure}

\begin{figure}
 \begin{center}
  \includegraphics[width=13cm]{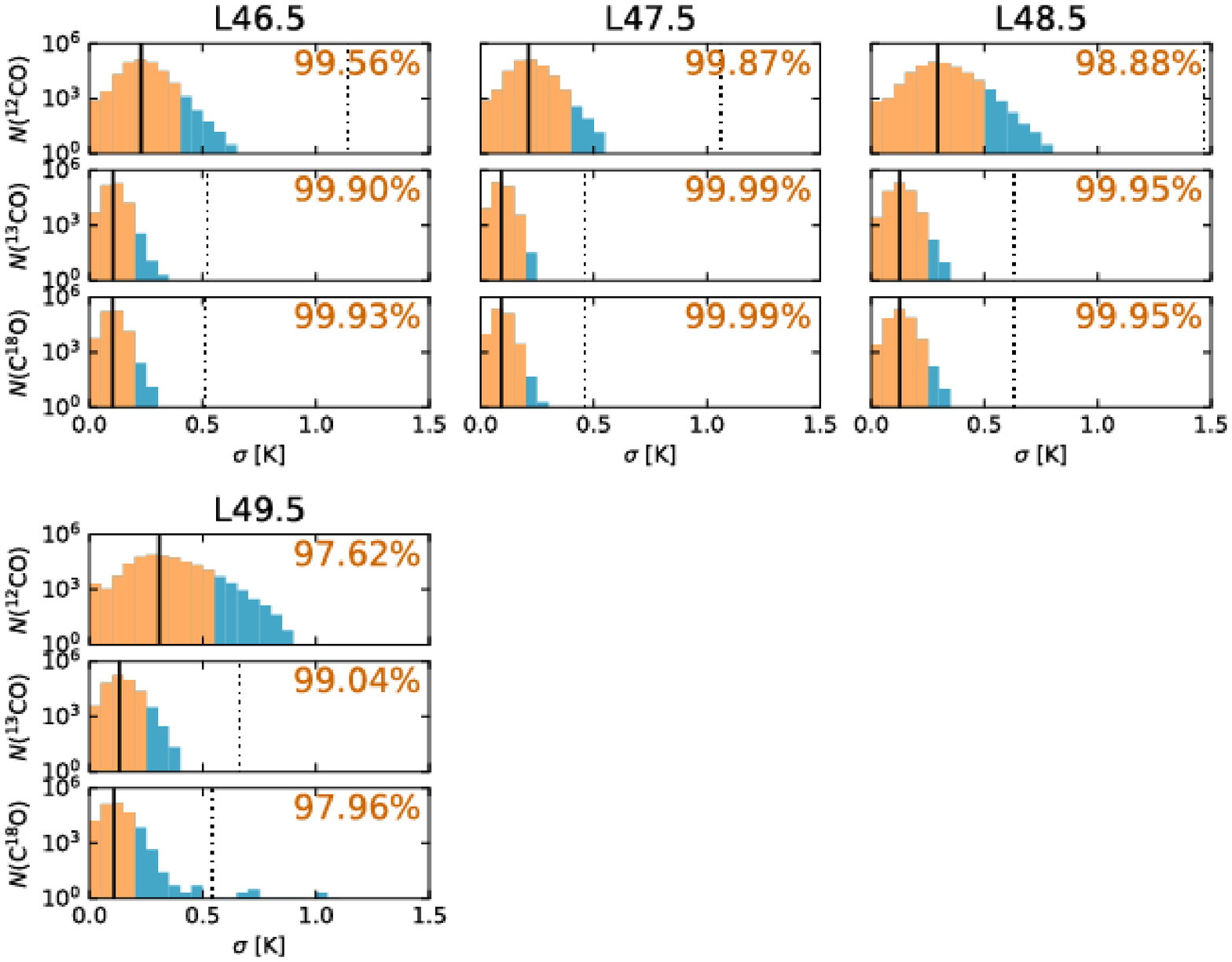}
 \end{center}
 \caption{Continued.}\label{fig:sighis5}
\end{figure}

\begin{figure}
 \begin{center}
  \includegraphics[width=13cm]{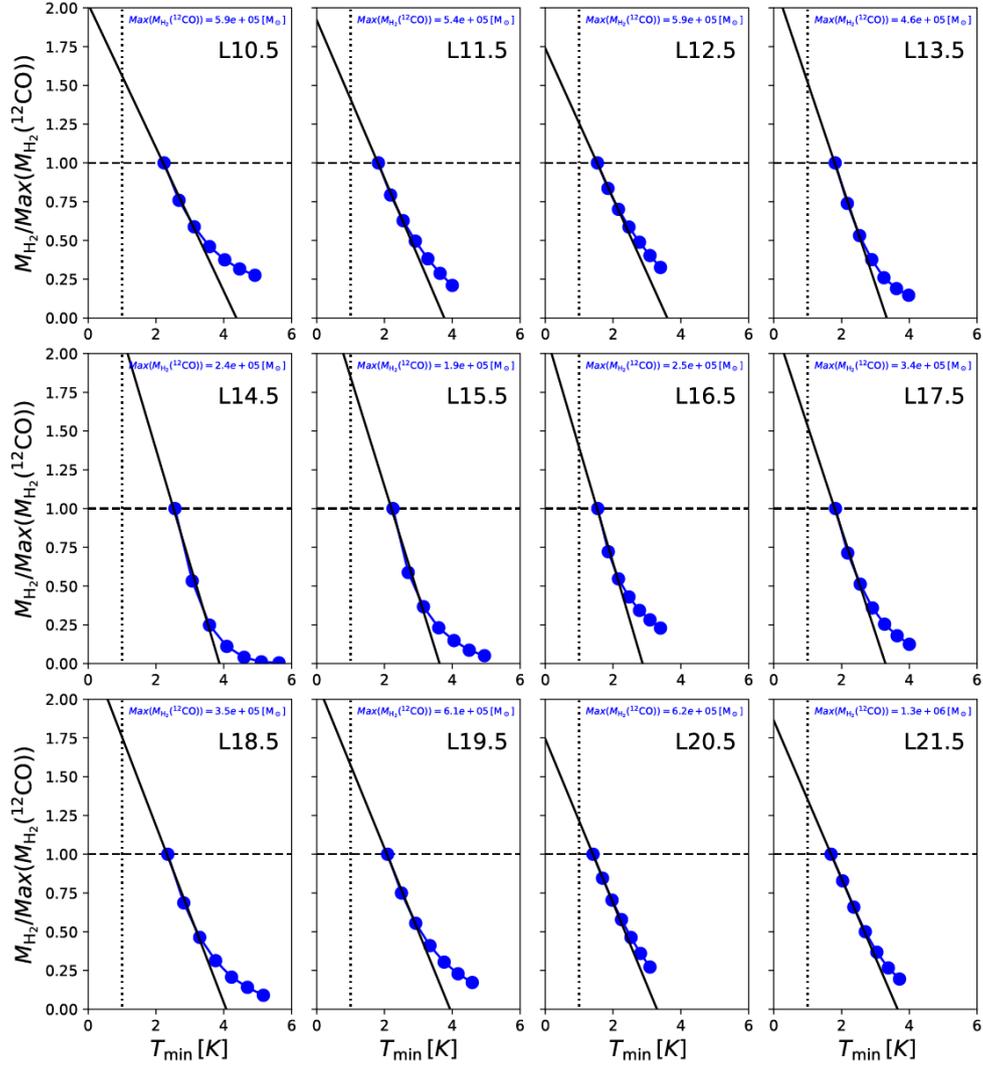}
 \end{center}
 \caption{Results of the extrapolations to derive the $\mtwo$ at $T_{\rm min} = 1$\,K. The blue circles indicates the normalized values of $\mtwo$ measured at $T_{\rm min} = 3, 4, 5, 6, 7, 8, 9, 10,$ and  $11 \times \sigma_{\rm med}$. The normalizations are made by dividing the $\mtwo$ by $max(M_{\rm H_2}{\rm (^{12}CO)})$, which is the maximum $\mtwo$. }\label{fig:mplot1}
\end{figure}

\begin{figure}
 \begin{center}
  \includegraphics[width=13cm]{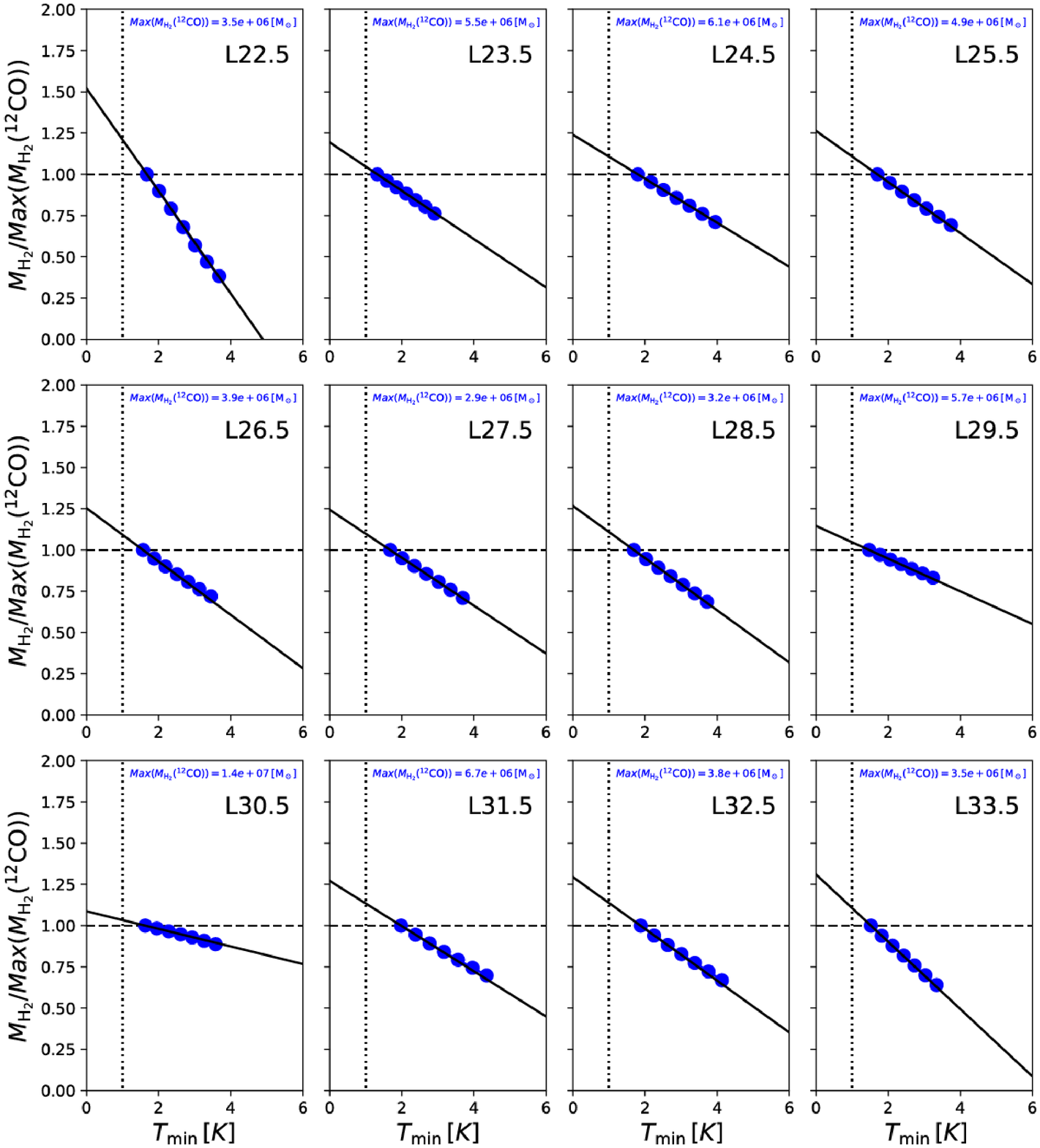}
 \end{center}
 \caption{Continued.}\label{fig:mplot2}
\end{figure}

\begin{figure}
 \begin{center}
  \includegraphics[width=13cm]{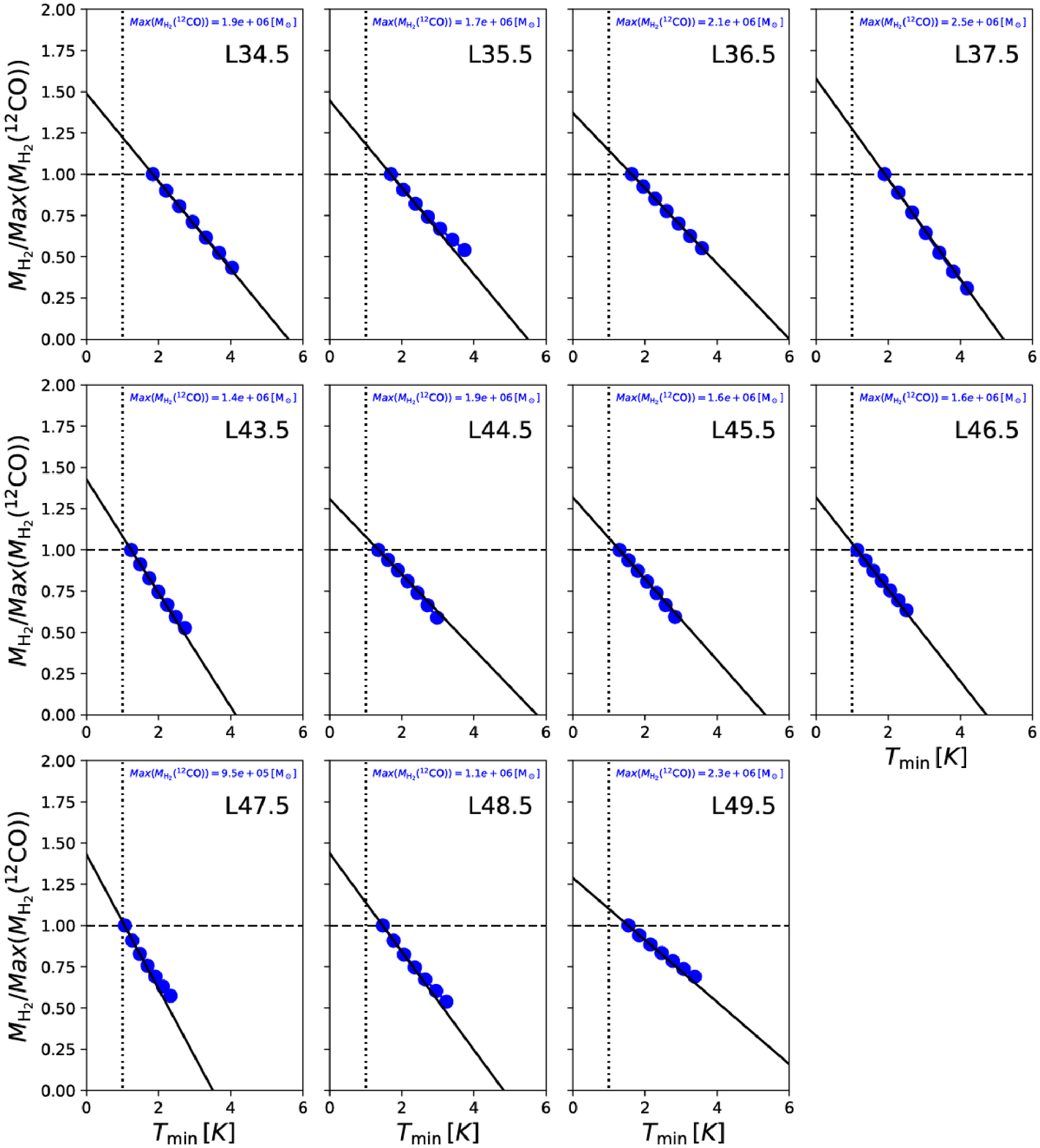}
 \end{center}
 \caption{Continued.}\label{fig:mplot3}
\end{figure}

\end{document}